\documentclass[journal=jctcce,manuscript=article]{achemso}

\pdfoutput=1

\usepackage[latin1]{inputenc}    % Silbentrennung und Eingabe von Umlauten.
\usepackage[T1]{fontenc}         % ermöglicht direkte Umlaute im Text. In die
\usepackage{times}               % nicer default font
\usepackage[english]{babel}
\usepackage{units}
\usepackage{xspace}
\usepackage{booktabs}            % Typographic improved tables
\usepackage{dcolumn}             % In Tabellen Ausrichtung am Dezimalpunkt mit S, lädt auch array
\usepackage{amsmath}
\usepackage{amssymb}
\usepackage{pifont}              % For \checkmark
\usepackage{numprint}
\npthousandsep{,}
\npdecimalsign{.}
\usepackage{spreadtab}           % Rechnen in Tabellen
\STsetdecimalsep{.}              % Was soll von spreadtab als Dezimalpunkt benutzt werden?
\usepackage{graphicx}            % zum Einbinden von Grafiken
\def\pics{./}                    % Pfad zum Verzeichnis der Abbildungen
\usepackage[official]{eurosym}   % Wir bezahlen bar und in Euro!
\usepackage{acronym}             % makes sure that acronyms are spelled out at least once 
\usepackage{numprint}
\usepackage[table]{xcolor}       % Farben, auch in Tabellen
\newcommand{\gromacs}{GROMACS\xspace}
\newcommand{\gromacsc}{GROMACS\cite{Hess:2008tf,Pronk:2013ef}\xspace}
\newcommand{\gromacsv}{\gromacs 4.6\xspace}
\newcommand{\QuotMarks}[1]{`#1'} % Quotation marks
\newcommand{\nsday}{\nicefrac{ns}{d}\xspace}

\newcommand{\fs}{\mbox{$^+$}\xspace}
\newcommand{\FS}{\mbox{$^\ddagger$}\xspace}

\newcommand{\mal}{\mbox{$\times$}\xspace}
\newcommand{\PreisEuro}[1]{\unit[\numprint{#1}]{\euro}}

\newcommand{\percent}{\%\xspace}
\newcommand{\degreeC}{$^\circ$C}

\newcommand{\cmark}{\ding{51}}
\newcommand{\xmark}{\ding{55}}
\newcommand{\nodlb}{\xmark}      % -dlb no option yields best performance
\newcommand{\dlb}{\cmark}        % -dlb yes option yields best performance
\newcommand{\odlbyes}{{\color{gray!50}(\cmark)}}  % only -dlb yes was tested
\newcommand{\hy}{ht\xspace}
\newcommand{\HY}{{\color{gray!50}(ht)}\xspace}

\newcommand{\nrank}{\mbox{$N_\text{rank}$}\xspace}
\newcommand{\nthread}{\mbox{$N_\text{th}$}\xspace}
\newcommand{\nth}{\mult{\nthread}}
\newcommand{\npme}{\mbox{$N_\text{PME}$}\xspace}
\newcommand{\ncore}{\mbox{$N_\text{c}$}\xspace}
\newcommand{\ncpu}{\mbox{$N_\text{CPU}$}\xspace}

\newcommand{\type}[1]{\texttt{\small #1}}

\newcommand{\eg}{e.\,g.\xspace}
\newcommand{\Eg}{E.\,g.\xspace}
\newcommand{\ie}{i.\,e.\xspace}

\newcommand{\two}{\mbox{$\times$2}\xspace}
\newcommand{\three}{\mbox{$\times$3}\xspace}
\newcommand{\four}{\mbox{$\times$4}\xspace}

\newcommand{\mult}{\multicolumn{1}{l}}

\newcolumntype{L}{>{$}l<{$}}
\newcolumntype{C}{>{$}c<{$}}
\newcolumntype{R}{>{$}r<{$}}
\newcolumntype{S}{D{.}{.}{-1}}
\newcolumntype{P}{ >{\cellcolor{gray!20}} D{.}{.}{-1}}
\newcolumntype{Z}[1]{>{\centering\arraybackslash $}p{#1} <{$}}  % zentriert mit Breitenangabe
\newcolumntype{g}{>{\cellcolor{gray!20}} c <{}}

\newcommand{\nanos}{\mult{(\nsday)}}
\newcommand{\euron}{\mult{{(\euro\ net)}}}
\newcommand{\DDGrid}{\multicolumn{3}{c}{\acs{DD} grid}}

\newcommand{\pcixcomment}{these nodes cannot use the full QDR IB bandwidth 
due to insufficient number of \ac{PCIe} lanes, see p.\ \pageref{pag:intelE3nodes}.}

\newcommand{\MPIbpc}{Theoretical and Computational Biophysics, Max Planck
Institute for Biophysical Chemistry, Am Fassberg 11, 37077 G\"ottingen, Germany}
\newcommand{\KTH}{Theoretical and Computational Biophysics, KTH Royal Institute of Technology, 
17121 Stockholm, Sweden}

\author{Carsten Kutzner}
\affiliation{\MPIbpc}
\email{ckutzne@gwdg.de}

\author{Szil\'ard P\'all}
\affiliation{\KTH}

\author{Martin Fechner}
\affiliation{\MPIbpc}

\author{Ansgar Esztermann}
\affiliation{\MPIbpc}

\author{Bert L. de Groot}
\affiliation{\MPIbpc}

\author{Helmut Grubm\"uller}
\affiliation{\MPIbpc}

\newcommand{\ManuscriptTitle}{Best bang for your buck:
GPU nodes for \gromacs biomolecular simulations}

\include{./biblio.bib}

\newacro{MD}{molecular dynamics}
\newacro{MEM}{membrane protein}
\newacro{RIB}{ribosome}
\newacro{SP}{single precision}
\newacro{DP}{double precision}
\newacro{MP}{mixed precision}
\newacro{GPU}{graphics processing unit}
\newacro{ECC}{error checking and correction}
\newacro{PME}{particle-mesh Ewald\cite{Essmann:1995vj}}
\newacro{LJ-PME}{Lennard-Jones PME}
\newacro{PP}{particle-particle}

\newacro{MPMD}{multiple program multiple data}
\newacro{SPMD}{single program multiple data}
\newacro{SIMD}{single instruction multiple data}
\newacro{HT}{Hyper-Threading}
\newacro{DD}{domain decomposition}
\newacro{PCIe}{PCI Express}
\newacro{IB}{Infiniband}
\newacro{HPC}{high performance computing}
\newacro{DLB}{dynamic load balancing}
\newacro{NUMA}{non-uniform memory access}
\title{\ManuscriptTitle}

\usepackage{xr}
\begin{document}

\hyphenation{off-loaded off-load off-loading Ge-Force over-clocked}

\maketitle

\begin{abstract}
The molecular dynamics simulation package \gromacs runs efficiently on a wide variety of hardware
from commodity workstations to high performance computing clusters. 
Hardware features are well exploited with a combination of SIMD, 
multi-threading, and MPI-based SPMD\,/\,MPMD parallelism,
while GPUs can be used as accelerators to compute
interactions offloaded from the CPU. 
Here we evaluate which hardware produces
trajectories with \gromacs 4.6 or 5.0 in the most economical way.
We have assembled and benchmarked compute nodes with various CPU\,/\,GPU combinations
to identify optimal compositions in terms of
raw trajectory production rate, performance-to-price ratio, 
energy efficiency, and several other criteria.
Though hardware prices are naturally subject to 
trends and fluctuations, general tendencies are clearly visible.
Adding any type of GPU significantly boosts a node's simulation performance.
For inexpensive consumer-class GPUs this improvement equally
reflects in the performance-to-price ratio.
Although memory issues in consumer-class GPUs could pass unnoticed since these cards
do not support ECC memory, unreliable GPUs can be sorted out with 
memory checking tools.
Apart from the obvious determinants for cost-efficiency
like hardware expenses and raw performance,
the energy consumption of a node is a major cost factor.
Over the typical hardware lifetime until replacement of a few years,
the costs for electrical power and cooling 
can become larger than the costs of the hardware itself.
Taking that into account, nodes with a
well-balanced ratio of CPU and consumer-class GPU resources
produce the maximum amount of \gromacs trajectory over their lifetime.
\end{abstract}

\section{Introduction}
Many research groups in the field of \ac{MD} simulation
and also computing centers need to make decisions on how to set up their compute clusters 
for running the \ac{MD} codes. 
A rich variety of \ac{MD} simulation codes is available, among them
CHARMM,\cite{CHARMM2009} Amber,\cite{AMBER2013} Desmond,\cite{Bowers2006} LAMMPS,\cite{Brown2012449}
ACEMD,\cite{Harvey2009} NAMD,\cite{phillips2005scalable} and \gromacsc.
Here we focus on \gromacs, which is among the fastest ones,  
and provide a comprehensive test intended to identify optimal hardware
in terms of \ac{MD} trajectory production per investment. 

One of the main benefits of \gromacs is its bottom-up performance-oriented design 
aimed at highly efficient use of the underlying hardware.
Hand-tuned compute kernels allow utilizing the \ac{SIMD} vector units of most consumer and HPC processor platforms,
while OpenMP multi-threading and \gromacs' built-in thread-MPI library together with \ac{NUMA}-aware optimizations
allow for efficient intra-node parallelism.
By employing a neutral-territory domain decomposition implemented with MPI, a simulation can be distributed across 
multiple nodes of a cluster. 
Beginning with version 4.6, 
the compute-intensive calculation of short-range non-bonded forces 
can be offloaded to \acp{GPU},
while the CPU concurrently computes all remaining forces such as long-range
electrostatics, bonds, etc., and updates the particle positions.\cite{Pall:2013gb}
Additionally, through \ac{MPMD} task-decomposition the long-range electrostatics calculation can be offloaded to a 
separate set of MPI ranks for better parallel performance.
This multi-level heterogeneous parallelization has been shown to achieve strong
 scaling to as little as 100 particles per core, at the same time reaching high absolute 
 application performance on a wide range of homogeneous 
and heterogeneous hardware platforms.\cite{Pall:2015,Abraham2015} 

A lot of effort has been invested over the years in software optimization,
resulting in \gromacs being one of the fastest \ac{MD} software engines available today.\cite{Hess:2008tf,Shaw2014}
\gromacs runs on a wide range of hardware,
but some node configurations produce trajectories more economically than others.
In this study we ask:
What is the \QuotMarks{optimal} hardware to run \gromacs on
and how can optimal performance be obtained?

Using a set of representative biomolecular systems 
we determine the simulation performance for various hardware combinations,
with and without \ac{GPU} acceleration.
For each configuration we aim to determine the run parameters
with the highest performance at comparable numerical accuracy. 
Therefore, this study also serves as a reference on what performance to expect 
for a given hardware. Additionally, we provide the \gromacs input files for
own benchmarks and the settings that gave optimum performance for each of
the tested node types.

Depending on the projects at hand, 
every researcher will have a somewhat different definition of \QuotMarks{optimal,} 
but one or more of the following criteria C1\,--\,C5 will typically be involved: 
\begin{itemize}
\item[C1 --] the performance-to-price ratio,
\item[C2 --] the achievable single-node performance,
\item[C3 --] the parallel performance or the \QuotMarks{time-to-solution,}
\item[C4 --] the energy consumption or the \QuotMarks{energy-to-solution,}
\item[C5 --] rack space requirements.
\end{itemize}
If on a fixed total budget for hardware, electricity, and cooling, 
the key task is to choose the hardware that produces the largest amount of
\ac{MD} trajectory for the investment. 

Here we focus on the most suitable hardware for \gromacs \ac{MD} simulations.
Due to the domain-specific requirements of biomolecular \ac{MD} and in particular that of
algorithms and implementation employed by \gromacs, such hardware will likely not be the best choice for a general-purpose cluster that is intended to serve a broad range of applications. At the same time, it is often possible
to pick a middle-ground that provides good performance both for \gromacs and other applications.

In the next section we will describe the key determinants for \gromacs performance,
and how \gromacs settings can be tuned for optimum 
performance on any given hardware.
Using two prototypic \ac{MD} systems, 
we will then systematically derive the settings yielding optimal performance
for various hardware configurations. 
For some representative hardware setups we will measure the power consumption to 
estimate the total \ac{MD} trajectory production costs including electrical power and cooling. 
Finally, for situations where simulation speed is crucial,
we will look at highly parallel simulations for several node types in a cluster setting.

\section{Key determinants for \gromacs performance}
\gromacs automatically detects a node's hardware resources such as
CPU cores, hardware thread support, and compatible \acp{GPU}, at run time.
The main simulation tool, \type{mdrun}, makes an educated guess on how to
best distribute the computational work onto the available resources.
When executed on a single node using its integrated, low-overhead thread-MPI library,
built-in heuristics can determine essentially all launch configurations automatically, 
including number of threads, ranks, and \ac{GPU} to rank assignment, allowing
to omit some or all of these options.
We use the term \QuotMarks{rank} for both MPI processes and thread-MPI ranks here;
both have the same functionality, whereas
thread-MPI ranks can only be used within the same node.
Additionally we use the term \QuotMarks{threads} or \QuotMarks{threading} to refer to
OpenMP threads; each rank may thus comprise a group of threads. 
\type{mdrun} optimizes the thread layout
for data locality and reuse also managing its own thread affinity settings.
Default settings typically result in a fairly good simulation performance,
and especially in single-node runs and on nodes with a single CPU and \ac{GPU}
often optimal performance is reached without optimizing settings manually.
However, tuning a standard simulation setup with \ac{PME} electrostatics 
for optimum performance on a compute node with \emph{multiple} CPUs and \acp{GPU} or
on a cluster of such nodes typically requires optimization of simulation and launch parameters.
To do this it is important to understand the underlying load distribution and balancing mechanisms.\cite{escidoc:2037317} 
The control parameters of these allow optimizing for simulation speed, without compromising numerical accuracy.

\subsection*{Load distribution and balancing mechanisms}
\label{sec:settings}

\gromacs uses \ac{DD} to split up the simulation system into 
$N_\text{DD} = \text{DD}_x \times \text{DD}_y \times \text{DD}_z$
initially equally-sized domains
and each of these is assigned to an MPI rank.
If \ac{DLB} is active, the sizes of the \ac{DD} cells are continuously adjusted
during the simulation to balance any
uneven computational load between the domains.

In simulations using \ac{PME}, \ac{MPMD} parallelization allows dedicating a group of \npme ranks
to the calculation of the long-range (reciprocal space) part of the Coulomb interactions, 
while the short-range (direct space) part is computed on the remaining $N_\text{DD}$ ranks.
A particle-mesh evaluation is also supported for the long-range component of 
the Lennard-Jones potential with the \ac{LJ-PME} implementation available since the 5.0 release.
\cite{Wennberg:2013bx,Abraham2015}
The coarse task-decomposition based on \ac{MPMD} allows reducing the number of ranks involved in the
costly all-to-all communication during 3D-FFT needed by the \ac{PME} computation,
which greatly reduces the communication overhead.\cite{Hess:2008tf,escidoc:2037317}
For a large number of ranks \nrank $\gg 8$,
peak performance is therefore usually reached with an appropriate separation \nrank $= N_\text{DD} + \npme$.
The number \npme of separate \ac{PME} ranks can be conveniently determined 
with the \type{g\_tune\_pme} tool,\footnote{From version 5.0 the \type{tune\_pme} command of the \type{gmx} tool.}
which is distributed with \gromacs since version 4.5.

When a supported \ac{GPU} is detected, the short-range part of Coulomb 
and van der Waals interactions are automatically offloaded,
while the long-range part, as needed for \ac{PME} or \ac{LJ-PME}, 
as well as bonded interactions are computed on the CPU. For the \ac{PME} computation, 
a fine \ac{PME} grid in combination with a short Coulomb cutoff results in a 
numerical accuracy comparable to that of a coarse grid with a large cutoff. 
Therefore, by increasing short-range interaction cutoff while also increasing the \ac{PME} grid spacing,
\gromacs can gradually shift computational load between \ac{PP} and \ac{PME} computation when the two are
executed on different resources. This is implemented in form of an automated static load-balancing 
between CPU and \ac{GPU} or between \ac{PP} and \ac{PME} ranks, and it is
carried out during the initial few hundreds to thousands of simulation steps.

By default, the \gromacs heterogeneous parallelization uses one \ac{GPU} per \ac{DD} cell,
mapping each accelerator to a \ac{PP} rank. If explicit threading parameters are omitted,
it also automatically distributes the available CPU cores among ranks within a node by spawning
the correct number of threads per rank.
Both thread count and order takes into account multiple hardware threads per core with \ac{HT}.
Using fewer and larger domains with \ac{GPU} acceleration allows reducing overhead associated
to GPU offloading like CUDA runtime as well as kernel startup and tail overheads.%
\footnote{Kernel \QuotMarks{tail} is the final, typically imbalanced part of a kernel execution
across multiple compute units (of a GPU), where some units already ran out of work while others are still active.}
On the other hand, since the minimum domain size is limited by cutoff and constraint restrictions,
using larger domains
also ensures that both small systems and systems with long-range constraints can be simulated using many \acp{GPU}.
Often however, performance is improved by using multiple domains per \ac{GPU}. In particular,
with more CPUs (or \ac{NUMA} regions) than \acp{GPU} per node
and also with large-core count processors, it is beneficial to reduce the thread count per rank 
by assigning multiple, \QuotMarks{narrower} ranks to a single GPU.
This reduces multi-threading parallelization overheads,
and by localizing domain data reduces 
cache coherency overhead and inter-socket communication.
Additional benefits come from multiple ranks sharing a \ac{GPU} as both compute kernels and transfers
dispatched from each rank using the same \ac{GPU} can overlap in newer CUDA versions.

CPU-\ac{GPU} and domain-decomposition load balancing are triggered simultaneously at the beginning 
of the run, which can lead to unwanted interaction between the two. 
This can have a detrimental effect on the performance in cases where DD cells are, or as a result of \ac{DLB} become
close in size to the cutoff in any dimension. In such cases, especially with pronounced DD load imbalance,
\ac{DLB} will quickly scale domains in an attempt to remove imbalance reducing the
domain sizes in either of the $x$, $y$, or $z$ dimensions to a value close to the original buffered cutoff.
This will limit the CPU-\ac{GPU} load-balancing in its ability to scale the cutoff, often
preventing it from shifting more work off of the CPU and leaving the \acp{GPU} under-utilized.
Ongoing work aims to eliminate the detrimental effect of this load balancer interplay with
a solution planned for the next \gromacs release.

Another scenario, not specific to \ac{GPU} acceleration, is where \ac{DLB} may indirectly reduce performance by
enforcing decomposition in an additional dimension. With \ac{DLB} enabled the \ac{DD} needs to account for domain resizing
when deciding on the number of dimensions required by the \ac{DD} grid. Without \ac{DLB} the same number of domains may be obtained by decomposing in fewer dimensions.
Although decomposition in all three dimensions
is generally possible, it is desirable to limit the number of dimensions
in order to reduce the volumes communicated.
In such cases, it can be faster to switch off \ac{DLB},
to fully benefit from \ac{GPU} offloading. 

\label{sec:noDlbOnGpu}

\subsection{Making optimal use of \acp{GPU}}
\label{sec:applicationClocks}
In addition to the load distribution and balancing mechanisms directly
controlled by \gromacs, with certain \ac{GPU} boards additional performance tweaks may be exploited.
NVIDIA Tesla cards starting with the GK110 micro-architecture
as well as some Quadro cards support a so-called \QuotMarks{application clock} setting.
This feature allows using a custom \ac{GPU} clock frequency either higher
or lower than the default value. Typically, this is used as a manual frequency boost to
trade available thermal headroom for improved performance, but
it can also be used to save power when lower \ac{GPU} performance is acceptable.
In contrast, consumer \acp{GPU} do not support application clocks but instead employ an automated
clock scaling (between the base and boost clocks published as part of the specs).
This can not be directly controlled by the user. 

\begin{figure}
\begin{center}
\includegraphics[width=7.5cm]{\pics 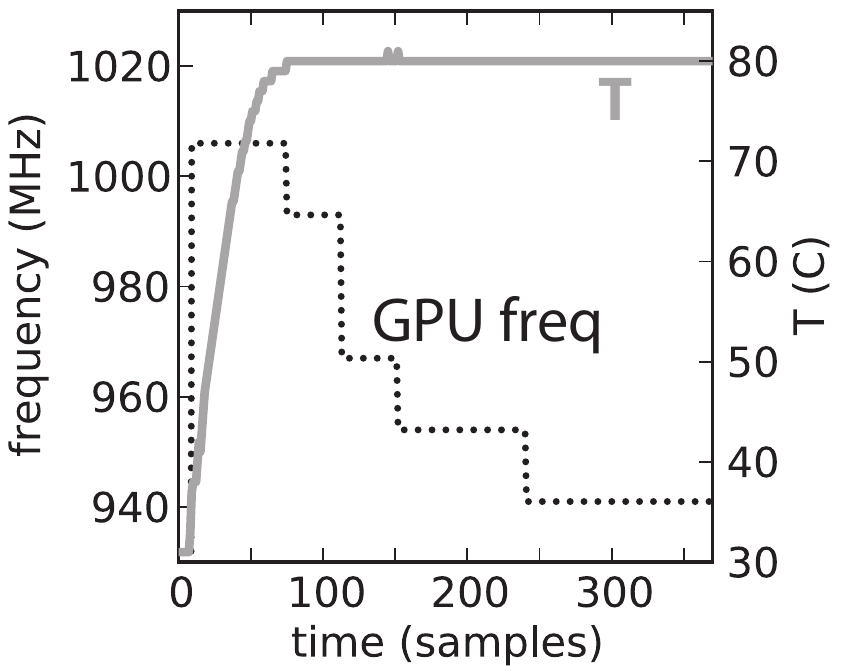}
\caption{Thermal throttling of the \ac{GPU} clock frequency on a GeForce GTX TITAN.
Starting from a cool, idle state at time $t=0$, at about $T=$~\unit[36]{\degreeC}, the \ac{GPU} is put under normal \gromacs load.
The clock frequency is first scaled to \unit[\numprint{1006}]{MHz}, but with the 
temperature quickly increasing due to the fan speed being capped at the default $60\%$,
the \ac{GPU} quickly reaches $T=$~\unit[80]{\degreeC}, starts throttling, gradually slowing down to \unit[941]{MHz}.}
\label{fig:freqVsTemp}
\end{center}
\end{figure}

A substantial thermal headroom can be available with compute applications because parts of the \ac{GPU}
board are frequently left under- or unutilized. Graphics-oriented functional units,
part of the on-board GDDR memory, and even the arithmetic units may idle,
especially in case of applications relying on heterogeneous parallelization.
In \gromacs the CPU-\ac{GPU} concurrent execution is possible only during force computation, and the \ac{GPU} is idle most of the time
outside this region, typically for 15\,--\,40\% of a time step.
This leaves enough thermal headroom to allow setting the highest application clock on all
\acp{GPU} to date (see Figure~\ref{fig:clocks} on p.~\pageref{pag:clocks}).

Increasing the \ac{GPU} core clock rate yields a proportional increase in non-bonded 
kernel performance. This will generally translate into improved \gromacs performance,
but its magnitude depends on how \ac{GPU}-bound the specific simulation is.
The expected performance gain is highest in strongly \ac{GPU}-bound cases
(where the CPU waits for results from GPU).
Here, the reduction in GPU kernel time translates into reduction in CPU wait time
hence improved application performance.
In balanced or CPU-bound cases the effective performance gain will often be smaller
and will depend on how well can the CPU-\ac{GPU} load-balancing make use of the increased GPU performance.
Note that there is no risk involved in using application clocks; 
even if a certain workload could generate high enough \ac{GPU} load for the
chip to reach its temperature or power limit, automated frequency throttling will ensure that the
limits will not be crossed.
The upcoming \gromacs 
version 5.1 will have built-in support for checking and setting the application clock of compute cards at runtime.

Indeed, frequency throttling is more common in case of the consumer boards, and
factory-over\-clocked parts can be especially prone to overheating.
Even standard clocked
desktop-oriented GeForce and Quadro cards come with certain disadvantages for compute use.
Being optimized for acoustics, desktop \acp{GPU} have their fan limited to approximately $60\%$ of the
maximum rotation speed. As a result, frequency throttling will occur as soon as 
the \ac{GPU} reaches its temperature limit, while the fan is kept at $\le 60\%$.
As illustrated on Figure~\ref{fig:freqVsTemp}, a GeForce GTX TITAN board installed in a well-cooled
rack-mounted chassis under normal \gromacs workload starts throttling already after a couple of minutes, 
successively dropping its clock speed by a total of 7\% in this case. This behavior is not uncommon and can cause
load-balancing issues and application slowdown as large as the \ac{GPU} slowdown itself.
The supporting information shows how to force the GPU fan speed to higher value.

Another feature, available only with Tesla cards, is the CUDA multi-process server (MPS) which provides two possible performance benefits.
The direct benefit is that it enables the overlap of tasks (both kernels and transfers)
issued from different MPI ranks to the same \ac{GPU}. As a result, the aggregate time to complete all 
tasks the assigned ranks will decrease. \Eg in a parallel run
with ~2000 atoms per MPI rank, using 6 ranks and a Tesla K40 GPU with CUDA MPS enabled we measured 30\%
reduction in the total GPU execution time compared to running without MPS.
A secondary, indirect benefit is that in some cases the CPU-side overhead of the CUDA runtime can be greatly reduced when,
instead of the pthreads-based thread-MPI, MPI processes are used (in conjunction with CUDA MPS to allow task overlap). 
Although CUDA MPS is not completely overhead-free, at high iteration rates of \unit[$<1$]{ms/step} quite common for \gromacs,
the task launch latency of the CUDA runtime causes up to $10-30\%$ overhead, but this can decreased 
substantially with MPI and MPS. In our previous example using 6-way GPU sharing 
the measured CUDA runtime overhead was reduced from 16\% to 4\%.

\section{Methods}
We will start this section by outlining the \ac{MD} systems used for performance evaluation. 
We will give details about the employed hardware 
and about the software environment in which the tests were done.
Then we will describe our benchmarking approach.

\subsection{Benchmark input systems}
\begin{table}
\begin{minipage}{\textwidth}
\renewcommand{\footnoterule}{} %
\renewcommand{\thefootnote}{\alph{footnote}}
\begin{center}
\caption{Specifications of the two \ac{MD} systems used for benchmarking.
}
\label{tab:systems}
\begin{tabular}{l c c} \toprule
\ac{MD} system                      & \acf{MEM}                   & \acf{RIB}                    \\
symbol used in plots                & $\bullet$                   & $\bigstar$                   \\ \midrule
\# particles                        & 81,743                      & 2,136,412                    \\
system size (nm)                    & 10.8$\times$10.2$\times$9.6 & 31.2$\times$31.2$\times$31.2 \\
time step length (fs)               & 2                           & 4                            \\
cutoff radii\footnotemark[1] (nm)            & 1.0                & 1.0                          \\
\acs{PME} grid spacing\footnotemark[1] (nm)  & 0.120              & 0.135                        \\
neighborlist update freq. CPU       & 10                          & 25                           \\
neighborlist update freq. \acs{GPU} & 40                          & 40                           \\
load balancing time steps           &  5,000 -- 10,000            & 1,000 -- 5,000               \\
benchmark time steps                &  5,000                      & 1,000 -- 5,000               \\
\bottomrule
\end{tabular}
\end{center}
\footnotetext[1]{Table lists the initial values of Coulomb cutoff and \acs{PME} grid spacing.
These are adjusted for optimal load balance at the beginning of a simulation.}
\end{minipage}
\end{table}
We used two representative biomolecular systems for benchmarking as summarized in Table~\ref{tab:systems}.
The \acs{MEM} system is a membrane channel protein embedded in a lipid bilayer 
surrounded by water and ions. 
With its size of $\approx$\,\unit[80]{k} atoms it serves as a prototypic example 
for a large class of setups used to study all kinds of membrane-embedded proteins.
\acs{RIB} is a bacterial ribosome in water with ions\cite{escidoc:1854600} and
with more than two million atoms an example of a rather large \ac{MD} system
that is usually run in parallel across several nodes.

\subsection{Software environment}
The benchmarks have been carried out with the most recent version of \gromacsv
available at the time of testing (see 5th column of Table~\ref{tab:environment}).  
Results obtained with version 4.6 will in the majority of cases hold for version 5.0 
since the performance of CPU and \ac{GPU} compute kernels have not changed substantially. 
Moreover, as long as compute kernel, threading and heterogeneous parallelization design 
remains largely unchanged, performance characteristics and optimization techniques described here
will translate to future versions, too.\footnote{This is has been verified for version 5.1
which is in beta phase at the time of writing.}

\begin{table}[t b]
\begin{minipage}{\textwidth}
\renewcommand{\footnoterule}{} %
\renewcommand{\thefootnote}{\alph{footnote}}
\begin{center}
\caption{Overview of the tested node types with the used software stack.
These nodes were combined with various \acp{GPU} from Table~\ref{tab:NVIDIA}.}
\label{tab:environment}
\begin{tabular}{l c c l l l l l}
\toprule
\multicolumn{4}{l}{Hardware per node}         & \multicolumn{3}{l}{Software stack (versions)}           \\ \cmidrule(r){1-4} \cmidrule(l){5-8}
Processor(s)              & total & RAM  & IB net- & GRO-     &         & MPI\footnotemark[1] &         \\
                          & cores & (GB) & work    & MACS     & GCC     & library             & \hspace{-3ex}CUDA\\ \midrule
Intel Core i7-4770K       &    4  &   8  &  --     &  4.6.7   & 4.8.3   &  --                 & 6.0     \\
Intel Core i7-5820K       &    6  &  16  &  --     &  4.6.7   & 4.8.3   &  --                 & 6.0     \\
Intel Xeon E3-1270v2      &    4  &  16  & QDR     &  4.6.5   & 4.4.7   & Intel 4.1.3         & 6.0     \\ %
Intel Xeon E5-1620        &    4  &  16  & QDR     &  4.6.5   & 4.4.7   & Intel 4.1.3         & 6.0     \\
Intel Xeon E5-1650        &    6  &  16  & QDR     &  4.6.5   & 4.4.7   & Intel 4.1.3         & 6.0     \\
Intel Xeon E5-2670        &    8  &  16  & QDR     &  4.6.5   & 4.4.7   & Intel 4.1.3         & 6.0     \\
Intel Xeon E5-2670v2      &   10  &  32  & QDR     &  4.6.5   & 4.4.7   & Intel 4.1.3         & 6.0     \\
Intel Xeon E5-2670v2 \two &   20  &  32  & QDR     &  4.6.5   & 4.4.7   & Intel 4.1.3         & 6.0     \\
Intel Xeon E5-2680v2 \two &   20  &  64  & FDR-14  &  4.6.7   & 4.8.4   & IBM PE 1.3          & 6.5     \\ %
Intel Xeon E5-2680v2 \two &   20  &  64  & QDR     &  4.6.7   & 4.8.3   & Intel 4.1.3         & 6.0     \\
AMD Opteron 6380 \two     &   32  &  32  & --      &  4.6.7   & 4.8.3   & --                  & 6.0     \\
AMD Opteron 6272 \four    &   64  &  32  & --      &  4.6.7   & 4.8.3   & --                  & --      \\
\bottomrule
\end{tabular}
\end{center}
\footnotetext[1]{For benchmarks across multiple nodes. 
On single nodes, \gromacs' built-in thread-MPI library was used.}
\end{minipage}
\end{table}

If possible, the hardware was tested in the same software environment
by booting from a common software image; 
on external \ac{HPC} centers the provided software environment was used. 
Table~\ref{tab:environment} summarizes the hard- and software situation
for the various node types.
The operating system was Scientific Linux 6.4 in most cases with the exception of
the FDR-14 Infiniband-connected nodes that 
were running SuSE Linux Enterprise Server 11.

For the tests on single nodes, \gromacs was compiled with OpenMP threads and its built-in thread-MPI library,
whereas across multiple nodes Intel's or IBM's MPI library was used.
In all cases, FFTW 3.3.2 was used for computing fast Fourier transformations. 
This was compiled using \type{--enable-sse2} for best \gromacs performance.\footnote{Although FFTW supports 
the AVX instruction set, due to limitations in its kernel auto-tuning functionality, 
enabling AVX support deteriorates performance on the tested architectures.}
For compiling \gromacs, the best possible \ac{SIMD} vector instruction set implementation was chosen
for the CPU architecture in question, \ie 128-bit AVX with FMA4 and XOP on AMD and 256-bit AVX on Intel
processors.

\gromacs can be compiled in \ac{MP} or in \ac{DP}. 
\ac{DP} treats all variables with \acl{DP} accuracy, whereas
\ac{MP} uses \ac{SP} for most variables, 
as \eg\ the large arrays containing positions, forces, and velocities, 
but double precision for some critical components like accumulation buffers.
It was shown that \ac{MP} does not deteriorate energy conservation.\cite{Hess:2008tf}
Since it produces \unit[1.4\,--\,2]{$\times$} more trajectory 
in the same compute time, it is in most cases preferable over \ac{DP}.\cite{Gruber:2010}
Therefore, we used \ac{MP} for the benchmarking.

\subsection{\ac{GPU} acceleration}

\begin{table}
\begin{minipage}{\textwidth}
\renewcommand{\footnoterule}{} %
\renewcommand{\thefootnote}{\alph{footnote}}
\newcommand{\anm}{\footnotemark[1]}
\begin{center}
\caption{Some \ac{GPU} models that can be used by \gromacs.
The upper part of the table lists \ac{HPC}-class Tesla cards,
below are the consumer-class GeForce GTX cards. 
Checkmarks (\cmark) indicate which were benchmarked.
For the GTX 980 \acp{GPU}, cards by different manufacturers
differing in clock rate were benchmarked (\fs and \FS symbols).
}
\label{tab:NVIDIA}
\STautoround{0} %
\small
\begin{spreadtab}{{tabular}{l l N{4}{0} N{4}{0} c N{4}{0} N{4}{0} c l}} \toprule
@NVIDIA          &@ architec-     &@\mult{CUDA}  &@\mult{min--max clock} &            &            & @\mult{clock rate} &@\mult{memory} & \SThidecol & @\mult{SP throughput} & @\mult{{$\approx$ price}} &            \\
@model           &@ ture          &@\mult{cores} &@\mult{rate (MHz)}     &            &            & @\mult{(MHz)}      &@\mult{(GB)}   & @lt. Wiki  & @\mult{(Gflop/s)}     & @\mult{{(\euro) (net)}}   &            \\ \midrule
@Tesla K20X\anm  &@ Kepler GK110  &     2688     & \SThidecol 735        & \SThidecol & \SThidecol &  732               &        6      &   3950     & \STcopy{v}{c3*g3/500} &       2800                & @\cmark    \\ %
@Tesla K40\anm   &@ Kepler GK110  &     2880     &            745        & @--        &  745       &  745               &       12      &   4290     &                       &       3100                & @\cmark    \\ \midrule
@GTX 680         &@ Kepler GK104  &     1536     &           1006        & @--        & 1058       & 1058               &        2      &   3090     &                       &       300                 & @\cmark    \\ %
@GTX 770         &@ Kepler GK104  &     1536     &           1046        & @--        & 1085       & 1110               &        2      &   3213     &                       &       320                 & @\cmark    \\ %
@GTX 780         &@ Kepler GK110  &     2304     &            863        & @--        &  900       &  902               &        3      &   3977     &                       &       390                 & @\cmark    \\ %
@GTX 780Ti       &@ Kepler GK110  &     2880     &            875        & @--        &  928       &  928               &        3      &   5046     &                       &       520                 & @\cmark    \\ %
@GTX TITAN       &@ Kepler GK110  &     2688     &            837        & @--        &  876       &  928               &        6      &   4500     &                       &       750                 & @\cmark    \\ %
@GTX TITAN X     &@ Maxwell GM200 &     3072     &           1002        & @--        & 1075       & 1002               &       12      &   6604     &                       &        @                  & @--        \\ %
@Quadro M6000    &@Maxwell GM200GL&     3072     &                       & @--        &            &  988               &       12      &            &                       &        @                  & @Fig.~\ref{fig:throughput}    \\
@GTX 970         &@ Maxwell GM204 &     1664     &           1050        & @--        & 1178       & 1050               &        4      &   3494     &                       &       250                 & @--        \\ %
@GTX 980         &@ Maxwell GM204 &     2048     &           1126        & @--        & 1216       & 1126               &        4      &   4612     &                       &       430                 & @\cmark    \\ %
@GTX 980\fs      &@ Maxwell GM204 &     2048     &           1126        & @--        & 1216       & 1266               &        4      &   4612     &                       &       450                 & @\cmark    \\ %
@GTX 980\FS      &@ Maxwell GM204 &     2048     &           1126        & @--        & 1216       & 1304               &        4      &   4612     &                       &       450                 & @\cmark    \\ %
\bottomrule
\end{spreadtab}
\end{center}
\footnotetext[1]{See Figure~\ref{fig:clocks} for how performance varies with clock rate of the Tesla cards,
all other benchmarks have been done with the base clock rates reported in this table.}
\end{minipage}
\end{table}

\gromacsv and later supports CUDA-compatible \acp{GPU} with compute capability 2.0 or higher.
Table~\ref{tab:NVIDIA} lists a selection of modern \acp{GPU} 
(of which all but the GTX 970 were benchmarked)
including some relevant technical information. 
The \ac{SP} column shows the \ac{GPU}'s maximum theoretical \ac{SP} flop rate,
calculated from the base clock rate (as reported by NVIDIA's \type{deviceQuery} program)
times the number of cores times two floating-point operations per core and cycle.
\gromacs exclusively uses \acl{SP} floating point (and integer) arithmetic on \acp{GPU}
and can therefore only be used in \acl{MP} mode with \acp{GPU}.
Note that at comparable theoretical \ac{SP} flop rate the Maxwell GM204 cards yield
a higher effective performance than Kepler generation cards due to better instruction scheduling and
reduced instruction latencies.

Since the \gromacs CUDA non-bonded kernels are by design strongly compute-bound,\cite{Pall:2013gb}
\ac{GPU} main memory performance has little impact on their performance. Hence,
peak performance of the \ac{GPU} kernels
can be estimated and compared within an architectural generation
simply from the the product of clock rate \mal cores.
\ac{SP} throughput is calculated from the base clock rate, but
the effective performance will greatly depend on the actual sustained
frequency a card will run at, which can be much higher. At the same time,
frequency throttling can lead to performance degradation as illustrated in Figure~\ref{fig:freqVsTemp}.

The price column gives an \emph{approximate} net original price of these \acp{GPU} as of 2014.
In general, cards with higher processing power (Gflop/s) are more expensive, 
however the TITAN and Tesla models have a significantly higher price due to 
their higher \ac{DP} processing power 
(\unit[\numprint{1310}]{Gflop/s} in contrast to at most \unit[210]{Gflop/s} for the 780Ti)
and their larger memory. 
Note that unless an \ac{MD} system is exceptionally large, 
or many copies are run simultaneously, 
the extra memory will almost never be used. For \ac{MEM},
\unit[$\approx$\,50]{MB} of \ac{GPU} memory is needed, 
and for \ac{RIB} \unit[$\approx$\,225]{MB}.
Even an especially large \ac{MD} system consisting of $\approx$\,\unit[12.5]{M} atoms
uses just about \unit[\numprint{1200}]{MB} and does therefore still
fit in the memory of any of the \acp{GPU} found in Table~\ref{tab:NVIDIA}.

\subsection{Benchmarking procedure}
The benchmarks were run for \numprint{2000}\,--\,\numprint{15000} steps, 
which translates to a couple of minutes wall clock runtime for the single-node benchmarks.
Balancing the computational load takes \type{mdrun} up to a few thousand time steps 
at the beginning of a simulation.
Since during that phase the performance is neither stable nor optimal, 
we excluded the first \numprint{1000}\,--\,\numprint{10000} steps from measurements using the
\type{-resetstep} or \type{-resethway} command line switches.
On nodes without a \ac{GPU}, we always activated \ac{DLB},
since the benefits of a balanced computational load between CPU cores usually outweigh the small
overhead of performing the balancing 
(see for example Figure~\ref{fig:threads}, black lines).
On \ac{GPU} nodes the situation is not so clear due to the 
competition between \ac{DD} and CPU-\ac{GPU} load balancing
mentioned in Section~\ref{sec:settings}. We therefore tested both with and 
without \ac{DLB} in most of the \ac{GPU} benchmarks.
All reported \ac{MEM} and \ac{RIB} performances are the average of two runs each,
with standard deviations on the order of a few percent 
(see Figure~\ref{fig:clocks} for an example of how the data scatter).

\subsubsection{Determining the single-node performance}
We aimed to find the optimal command-line settings for each hardware configuration
by testing the various parameter combinations as mentioned in Section~\ref{sec:settings}.
On individual nodes with \ncore cores, %
we tested the following settings using thread-MPI ranks:
\begin{itemize}
\item[(a)] \nrank = \ncore
\item[(b)] a single process with \nthread = \ncore threads
\item[(c)] combinations of \nrank ranks with \nthread threads each, 
with \nrank \mal \nthread = \ncore (hybrid parallelization)
\item[(d)] for \ncore $\ge 20$ without \ac{GPU} acceleration, we additionally
checked with \type{g\_tune\_pme} whether separate ranks for the long-range \ac{PME} part do
improve performance
\end{itemize}
For most of the hardware combinations, we checked (a)\,--\,(d)
with and without \ac{HT}, if the processor supports it.
The supporting information contains a \type{bash} script that
automatically performs tests (a)\,--\,(c).

To share \acp{GPU} among multiple \ac{DD} ranks, current versions of \type{mdrun} require a 
custom \type{-gpu\_id} string specifying the mapping between \ac{PP} ranks
and numeric \ac{GPU} identifiers.
To obtain optimal launch parameters on \ac{GPU} nodes, we automated constructing
the \type{-gpu\_id} string based on the number of \ac{DD} ranks and \acp{GPU}
and provide the corresponding \type{bash} script in the supporting information.

\subsubsection{Determining the parallel performance}
To determine the optimal performance across many CPU-only nodes in parallel,
we ran \type{g\_tune\_pme} 
with different combinations of ranks and threads.
We started with as many ranks as cores \ncore in total (no threads), 
and then tested two or more threads per rank with an appropriately 
reduced number of ranks as in (c), with and without \ac{HT}. 

When using separate ranks for the direct and reciprocal space parts of \ac{PME}
($N = N_\text{DD} + \npme$) on a cluster of \ac{GPU} nodes,
only the $N_\text{DD}$ direct space ranks can make use of \acp{GPU}.
Setting whole nodes aside for the \ac{PME} mesh calculation
would mean leaving their \ac{GPU}(s) idle. 
\label{sec:homogeneous}
To prevent leaving resources unused with separate \ac{PME} ranks, 
we assigned as many direct space (and reciprocal space) ranks 
to each node as there are \acp{GPU} per node, resulting in a homogeneous,
interleaved \ac{PME} rank distribution.
On nodes with two \acp{GPU} each, \eg, we placed $N = 4$ ranks
($N_\text{DD} = \npme = 2$)
with as many threads as needed to make use of all available cores.
The number of threads per rank may even differ for $N_\text{DD}$ and \npme.
In fact, an uneven thread count can be used to balance the compute power
between the real and the reciprocal ranks.
On clusters of \ac{GPU} nodes, we tested all of the above scenarios (a)\,--\,(c)
and additionally checked whether a homogeneous, interleaved \ac{PME} rank distribution
improves performance.

\section{Results}
This section starts with four pilot surveys that 
assess \ac{GPU} memory reliability (i), and
evaluate the impact of compiler choice (ii),
neighbor searching frequency (iii), 
and parallelization settings (iv) on the \gromacs performance.
From the benchmark results and the hardware costs we will
then derive for various node types how much \ac{MD} trajectory is produced per invested \euro.
We will compare performances of nodes with and without \acp{GPU} and also
quantify the performance dependence on the \ac{GPU} application clock setting. 
We will consider the energy efficiency of several node types and show
that balanced CPU-\ac{GPU} resources are needed for a high efficiency.
We will show how running multiple simulations concurrently maximizes
throughput on \ac{GPU} nodes. Finally, we will examine the parallel
efficiency in strong scaling benchmarks for a selected subset of node types.

\subsection{\ac{GPU} error rates}
Opposed to the GeForce GTX consumer \acp{GPU}, 
the Tesla \ac{HPC} cards offer \acfi{ECC}\acused{ECC} memory.
\ac{ECC} memory, as also used in CPU server hardware, 
is able to detect and possibly correct random memory bit-flips that may rarely occur.
While in a worst-case scenario 
such events could lead to silent memory corruption and incorrect simulation results,
their frequency is extremely low.\cite{Shi:2009tv,walker2013investigation}
Prior to benchmarking, 
we performed extensive \ac{GPU} stress-tests on a total of 297 consumer-class
\acp{GPU} (Table~\ref{tab:errors}) using tools that test for \QuotMarks{soft errors} in the \ac{GPU}
memory subsystem and logic using a variety of proven test patterns.\cite{haque2010hard}
Our tests allocated the entire available \ac{GPU} memory and ran for $\geq$\,\numprint{4500}
iterations, corresponding to several hours of wall-clock time.
The vast majority of cards were error-free, %
but for 8 \acp{GPU}, errors were detected.
Individual error rates differed considerably from one card to another with
the largest rate observed for a 780Ti, where during 10,000 iterations
$>$\,50 Million errors were registered. Here, already the first iteration of the memory
checker picked up $>$\,1,000 errors. On the other end of the spectrum were
cards exhibiting only a couple of errors over 10,000 iterations, including the two 
problematic 980\fs. 
Error rates were close to constant for each of the four repeats over 10,000 iterations.
All cards with detected problems were replaced.
\begin{table}
\begin{center}
\caption{Frequency of consumer-class \acp{GPU} exhibiting memory errors.}
\label{tab:errors}
\begin{tabular}{l l R R c} \toprule
\mult{NVIDIA} & GPU memory        &\mbox{\# of cards} &\mbox{\# memtest}   &\mbox{\# cards}    \\
\mult{model}  & checker\cite{haque2010hard} &\mbox{tested} &\mbox{iterations} &\mbox{with errors}\\ \midrule
GTX 580       & \type{memtestG80} &    1              &    10,000          & --                \\
GTX 680       & \type{memtestG80} &    50             &     4,500          & --                \\
GTX 770       & \type{memtestG80} &    100            &     4,500          & --                \\
GTX 780       & \type{memtestCL}  &    1              &    50,000          & --                \\
GTX TITAN     & \type{memtestCL}  &    1              &    50,000          & --                \\
GTX 780Ti     & \type{memtestG80} &    70             & 4\times 10,000     & 6                 \\
GTX 980       & \type{memtestG80} &     4             & 4\times 10,000     & --                \\
GTX 980\fs    & \type{memtestG80} &    70             & 4\times 10,000     & 2                 \\
\bottomrule
\end{tabular}
\end{center}
\end{table}

\subsection{Impact of compiler choice}
The impact of the compiler version on the simulation performance is quantified
in Table~\ref{tab:compilers}.  
From all tested compilers, GCC 4.8
provides the fastest executable on both AMD and Intel platforms. 
On \ac{GPU} nodes, the difference between the fastest and slowest
executable is at most 4\percent, but without \acp{GPU} it can reach  
20\percent. Table~\ref{tab:compilers} can also be used to normalize benchmark
results obtained with different compilers.

\begin{table}
\begin{center}
\caption{\gromacs 4.6 single-node performance with thread-MPI (and CUDA 6.0) using different compiler versions on AMD and Intel hardware
with and without \acp{GPU}.
The last column shows the speedup compared to GCC 4.4.7 
calculated from the average of the speedups of the \ac{MEM} and \ac{RIB} benchmarks.}
\label{tab:compilers}
\newcommand{\xxx}{\hspace{3ex}}
\begin{spreadtab}{{tabular}{l l N{4}{1} c N{4}{3} c N{4}{1} }} \toprule
@ Hardware                   &@ Compiler    &@ \mult{MEM (\nsday)} &@ ratio                &@ \mult{RIB (\nsday)} &@ ratio                &@\mult{av. speedup (\%)}                     \\ \midrule
@ AMD 6380 \mal 2            &@ GCC  4.4.7  &  round(14.043, 1)    & \STcopy{v}{c2/!c!2}   &  round(0.986, 2)     & \STcopy{v}{e2/!e!2}   & \STcopy{v}{round(100*((d2+f2)/2 - 1), 1)}   \\
@                            &@ GCC  4.7.0  &  round(15.609, 1)    &                       &  round(1.106, 2)     &                       &                                             \\
@                            &@ GCC  4.8.3  &  round(16.025, 1)    & \SThidecol            &  round(1.135, 2)     & \SThidecol            &                                             \\
@                            &@ ICC 13.1    &  round(12.482, 1)    &                       &  round(0.959, 2)     &                       &                                             \\ \midrule
@ AMD 6380 \mal 2            &@ GCC  4.4.7  &  round(40.456, 1)    & \STcopy{v}{c6/!c!6}   &  round(3.044, 2)     & \STcopy{v}{e6/!e!6}   & \STcopy{v}{round(100*((d6+f6)/2 - 1), 1)}   \\
@ \xxx with 2\mal GTX 980\fs &@ GCC  4.7.0  &  round(38.8845,1)    &                       &  round(3.091, 2)     &                       &                                             \\
@                            &@ GCC  4.8.3  &  round(40.209, 1)    &                       &  round(3.1415,2)     &                       &                                             \\
@                            &@ ICC 13.1    &  round(39.651, 1)    &                       &  round(3.0885,2)     &                       &                                             \\ \midrule
@ Intel E5-2680v2 \mal 2     &@ GCC  4.4.7  &  round(21.606, 1)    & \STcopy{v}{c10/!c!10} &  round(1.628, 2)     & \STcopy{v}{e10/!e!10} & \STcopy{v}{round(100*((d10+f10)/2 - 1), 1)} \\
@                            &@ GCC  4.8.3  &  round(26.798 ,1)    &                       &  round(1.858, 2)     &                       &                                             \\
@                            &@ ICC 13.1    &  round(24.570 ,1)    &                       &  round(1.876, 2)     &                       &                                             \\
@                            &@ ICC 14.0.2  &  round(25.203 ,1)    &                       &  round(1.811, 2)     &                       &                                             \\ \midrule
@ Intel E5-2680v2 \mal 2     &@ GCC  4.4.7  &  round(61.2335,1)    & \STcopy{v}{c14/!c!14} &  round(4.4055,2)     & \STcopy{v}{e14/!e!14} & \STcopy{v}{round(100*((d14+f14)/2 - 1), 1)} \\
@ \xxx with 2\mal GTX 980\fs &@ GCC  4.8.3  &  round(62.341, 1)    &                       &  round(4.688, 2)     &                       &                                             \\
@                            &@ ICC 13.1    &  round(60.3  , 1)    &                       &  round(4.7835,2)     &                       &                                             \\ 
\bottomrule
\end{spreadtab}
\end{center}
\end{table}

\subsection{Impact of neighbor searching frequency}
With the advent of the Verlet cutoff scheme implementation in version 4.6,
the neighbor searching frequency has become a merely performance-related parameter.
This is enabled by the automated pair list buffer calculation 
based on the a maximum error tolerance and
a number of simulation parameters and properties of the simulated system including 
search frequency, temperature, atomic displacement distribution and the shape of the
potential at the cutoff.\cite{Pall:2013gb}

Adjusting this frequency allows trading the computational cost of 
searching for the computation of short-range forces.
As the \ac{GPU} is idle during list construction on the CPU, reducing the 
search frequency also increases the average CPU-\ac{GPU} overlap.
Especially in multi-\ac{GPU} runs where \ac{DD} is done at the same step 
as neighbor search, decreasing the search
frequency can have considerable performance impact.
\begin{figure}[h] 
\begin{center}
\includegraphics[width=9cm]{\pics 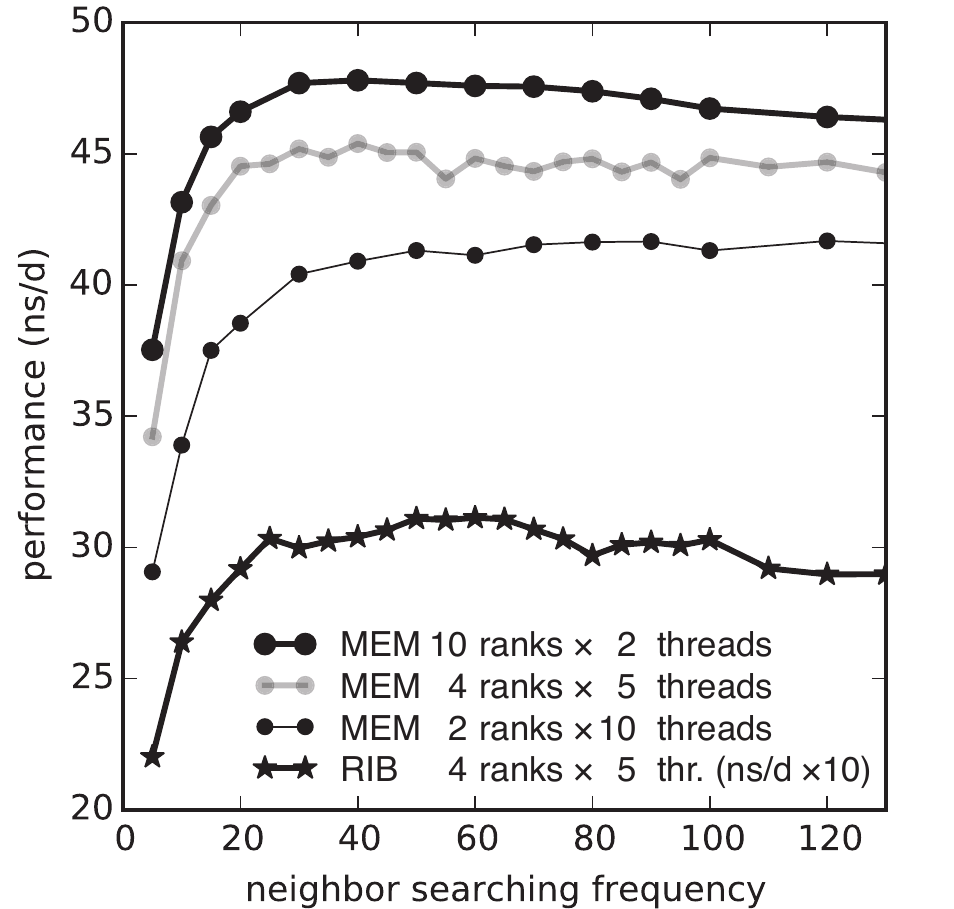}
\caption{
Impact of neighbor searching frequency on the performance 
on a node with 2\mal{}E5-2680v2 processors and 2\mal{}K20X \acp{GPU}.
In the \ac{MEM} benchmark the number of ranks and threads per rank was also varied.}
\label{fig:nstlist}
\end{center}
\end{figure}
Figure~\ref{fig:nstlist} 
indicates that the search frequency optimum is between 20 and 70 time steps.
The performance dependence is most pronounced for values $\le 20$, 
where performance quickly deteriorates.
In our benchmarks we used a value of 40 on \ac{GPU} nodes (see Table~\ref{tab:systems}).

\subsection{Influence of hybrid parallelization settings and \ac{DLB}}

\begin{figure}
\begin{center}
\includegraphics[width=16.5cm]{\pics 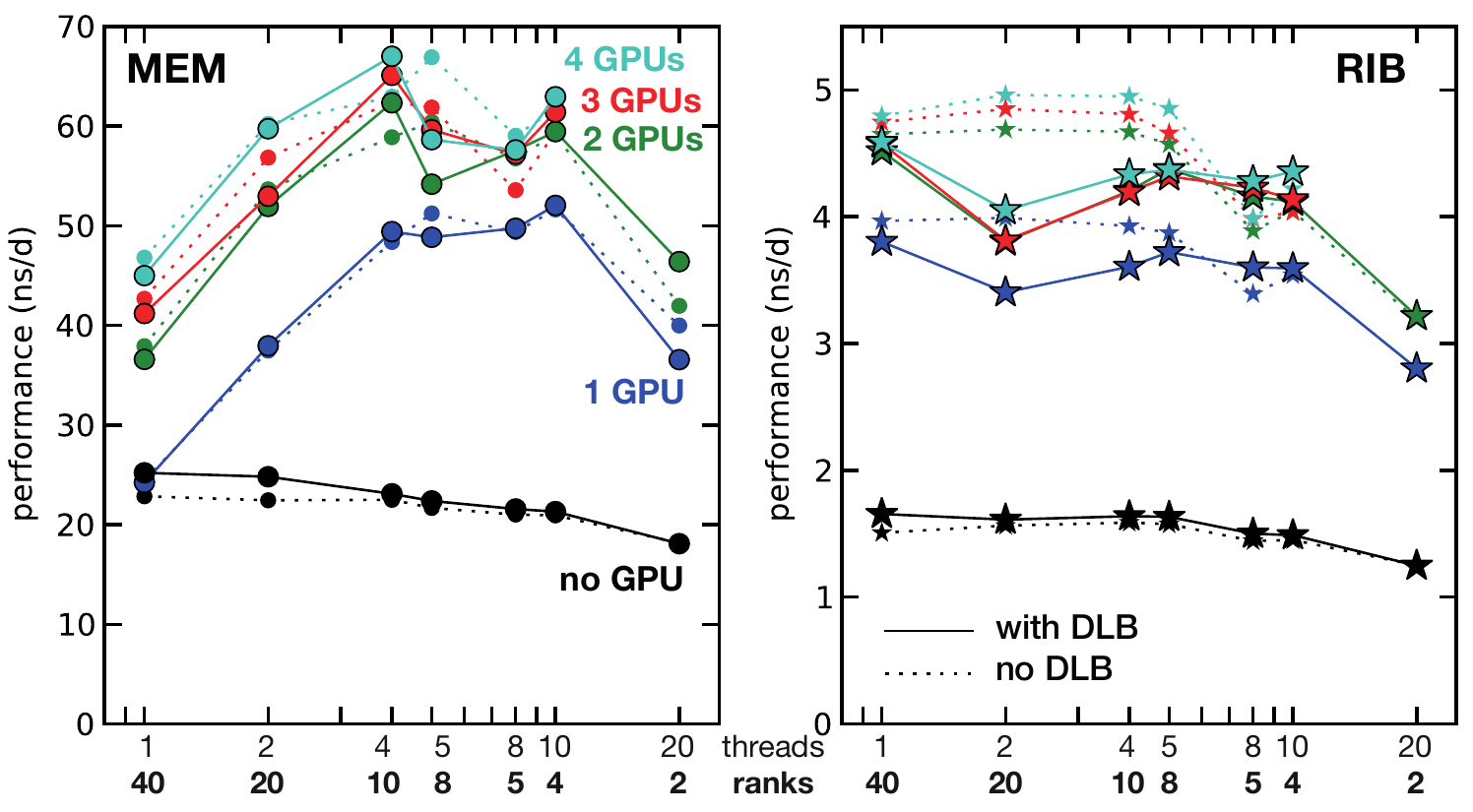}
\caption{
Single-node performance as a function of the number of \acp{GPU} (color coded) 
and of how the 40 hardware threads are exploited using a combination
of MPI ranks and OpenMP threads.
Solid lines show performance with, dotted lines without \ac{DLB}.
Test node had 2\mal E5-2680v2 processors and 4\mal GTX 980\fs~\acp{GPU}. 
Left panel \ac{MEM}, right panel \ac{RIB} benchmark system.
}
\label{fig:threads}
\end{center}
\end{figure}

The hybrid (OpenMP\,/\,MPI) parallelization approach in 
\gromacs can distribute computational work on the available CPU cores in various ways. 
Since the MPI and OpenMP code paths exhibit a different
parallel scaling behaviour,\cite{Pall:2015} 
the optimal mix of ranks and threads depends on the used hardware
and \ac{MD} system, as illustrated in Figure~\ref{fig:threads}.

For the CPU-only benchmarks shown in the figure (black lines),
pure MPI parallelization yields the highest performance,
which is often the case on nodes without \acp{GPU} (see
Fig.~4 in Ref.~\citep{Pall:2015}).
For multi-socket nodes with \ac{GPU}(s) and for nodes with multiple \acp{GPU},
the highest performance is usually reached with hybrid parallelism 
(with an optimum at about 4\,--\,5 threads per MPI rank, colored curves). 
The performance differences between the individual parallel
settings can be considerable:
for the single-\ac{GPU} setting of the \ac{MEM} system, \eg, 
choosing 40 MPI ranks results in less than half the performance of
the optimal settings, which are 4 MPI ranks and 10 threads each
(\unit[24]{\nsday} compared to \unit[52]{\nsday}, 
see blue line in Figure~\ref{fig:threads}).
The settings at the performance optimum are provided in the benchmark summary
Tables~\ref{tab:numbersAQP}, \ref{tab:numbersRIB}, \ref{tab:scalingAQP}, and \ref{tab:scalingRIB}.

As described in Section~\ref{sec:noDlbOnGpu}, especially with \acp{GPU}, 
\ac{DLB} may in some cases cause performance degradation. 
A prominent example is the right plot of Figure~\ref{fig:threads},
where the highest \ac{RIB} performances are recorded without \ac{DLB} when using \acp{GPU}.
However, there are also cases where the performance is similar
with and without \ac{DLB}, as \eg\ in the 4-\ac{GPU}
case of the left plot (light blue).

\subsection{Fitness of various node types}

\newcommand*{\GhZ}{\hspace{2ex}}
\newcommand*{\ghz}[1]{\GhZ{(#1 GHz)}}
\newcommand*{\GHZ}   {\GhZ{clock rate}}
\newcommand*{\myTableSettings}{\renewcommand{\baselinestretch}{0.80}\footnotesize}

\begin{table}
\caption{Single-node performance $P$ of the \ac{MEM} benchmark on various node types. 
U = rack space requirements in units per node, D for desktop chassis.
Prices do not include \ac{IB} network adapter.
\vspace{-7mm}}
\label{tab:numbersAQP}
\STautoround{1} %
\FPset\fAQP{205}
\begin{center}
\myTableSettings
\begin{spreadtab}{{tabular}{c l c l r r r c r c P N{5}{0} S}} \toprule
@ U &@ processor(s)    &@CPUs \mal  &@ \acsp{GPU}  &@ \DDGrid         &@\npme&@\nth&@\mult{\ac{DLB}} &@\mult{$P$}  &@\mult{{$\approx$\,cost}} & @\mult{\nsday per}        \\ 
@   &@\GHZ             &@cores      &@             &@x  &@y  &@z      &      &     &            &@\nanos           &@\euron           & @\mult{\fAQP\ \euro}              \\ \midrule
@ D &@ i7-4770K        &@ 1 \mal  4 &@  --         &  2 &  3 &  1     &  2   &  1  & @\odlbyes  & round( 7.364 ,1) &  800             & \STcopy{v}{[-2,0]/([-1,0]/\fAQP)} \\
    &@\ghz{3.4--3.9}   &            &@ 980\FS      &  1 &  1 &  1     & @--  &  8  & @--        & round(26.081 ,1) &  800+450         &                                   \\ \midrule
@ D &@ i7-5820K        &@ 1 \mal  6 &@  --         &  3 &  3 &  1     &  3   &  1  & @\dlb      & round(10.081, 1) &  850             &                                   \\ 
    &@\ghz{3.3--3.6}   &            &@ 770         &  1 &  1 &  1     & @--  & 12  & @--        & round(26.532, 1) &  850+320         &                                   \\
    &                  &            &@ 980\FS      &  1 &  1 &  1     & @--  & 12  & @--        & round(31.9955,1) &  850+540         &                                   \\ \midrule
  1 &@ E3-1270v2       &@ 1 \mal  4 &@  --         &  1 &  1 &  1     & @--  &  8  & @--        & round( 5.344 ,1) & 1080             &                                   \\ %
    &@\ghz{3.5}        &            &@ 680         &  1 &  1 &  1     & @--  &  8  & @--        & round(19.9675,1) & 1380             &                                   \\ 
    &                  &            &@ 770         &  1 &  1 &  1     & @--  &  8  & @--        & round(20.530 ,1) & 1400             &                                   \\ \midrule %
  1 &@ E5-1620         &@ 1 \mal  4 &@ 680         &  1 &  1 &  1     & @--  &  8  & @--        & round(21.0375,1) & 1900             &                                   \\ %
    &@\ghz{3.6--3.8}   &            &@ 770         &  1 &  1 &  1     & @--  &  8  & @--        & round(21.7   ,1) & 1900             &                                   \\ 
    &@                 &            &@ 780         &  1 &  1 &  1     & @--  &  8  & @--        & round(21.8   ,1) & 1970             &                                   \\ %
    &@                 &            &@ 780Ti       &  1 &  1 &  1     & @--  &  8  & @--        & round(23.4   ,1) & 1970+130         &                                   \\ %
    &@                 &            &@ TITAN       &  1 &  1 &  1     & @--  &  8  & @--        & round(23.8   ,1) & 1900+430         &                                   \\ \midrule %
  1 &@ E5-1650         &@ 1 \mal  6 &@ 680         &  1 &  1 &  1     & @--  & 12  & @--        & round(22.6   ,1) & 2170             &                                   \\ %
    &@\ghz{3.2--3.8}   &            &@ 770         &  1 &  1 &  1     & @--  & 12  & @--        & round(23.4   ,1) & 2170             &                                   \\ %
    &                  &            &@ 780         &  1 &  1 &  1     & @--  & 12  & @--        & round(25.0   ,1) & 2170+70          &                                   \\ %
    &                  &            &@ 780Ti       &  1 &  1 &  1     & @--  & 12  & @--        & round(27.0   ,1) & 2370             &                                   \\ %
    &                  &            &@ 680\two     &  2 &  1 &  1     & @--  &  6  & @\odlbyes  & round(24.8   ,1) & 2470             &                                   \\ %
    &                  &            &@ 770\two     &  2 &  1 &  1     & @--  &  6  & @\odlbyes  & round(25.1   ,1) & 2470             &                                   \\ \midrule %
  1 &@ E5-2670         &@ 1 \mal  8 &@ 770         &  1 &  1 &  1     & @--  & 16  & @--        & round(26.9   ,1) & 2800             &                                   \\ %
    &@\ghz{2.6--3.3}   &            &@ 780         &  1 &  1 &  1     & @--  & 16  & @--        & round(28.34  ,1) & 2800+70          &                                   \\ %
    &                  &            &@ 780Ti       &  1 &  1 &  1     & @--  & 16  & @--        & round(29.6   ,1) & 2800+200         &                                   \\ %
    &                  &            &@ TITAN       &  1 &  1 &  1     & @--  & 16  & @--        & round(29.30  ,1) & 2800+430         &                                   \\ %
    &                  &            &@ 770\two     &  2 &  1 &  1     & @--  &  8  & @\odlbyes  & round(27.6   ,1) & 3120             &                                   \\ \midrule %
  1 &@ E5-2670v2       &@ 1 \mal 10 &@ --          &  4 &  5 &  1     & @--  &  1  & @\dlb      & round(11.171 ,1) & 2800-320         &                                   \\
    &@\ghz{2.5--3.3}   &            &@ 770         &  1 &  5 &  1     & @--  &  4  & @\nodlb    & round(28.9575,1) & 2800             &                                   \\
    &                  &            &@ 780         &  1 &  1 &  1     & @--  & 20  & @--        & round(29.8   ,1) & 2800+70          &                                   \\
    &                  &            &@ 780Ti       &  1 &  5 &  1     & @--  &  4  & @\odlbyes  & round(31.4535,1) & 2800+200         &                                   \\
    &                  &            &@ TITAN       &  1 &  1 &  1     & @--  & 20  & @--        & round(32.7   ,1) & 2800+430         &                                   \\ 
    &                  &            &@ 980         &  1 &  5 &  1     & @--  &  4  & @\odlbyes  & round(33.6125,1) & 2800+100         &                                   \\
    &                  &            &@ 770\two     & 10 &  1 &  1     & @--  &  2  & @\dlb      & round(33.6835,1) & 3120             &                                   \\
    &                  &            &@ 780Ti\two   & 10 &  1 &  1     & @--  &  2  & @\odlbyes  & round(35.7205,1) & 3120+2*200       &                                   \\
    &                  &            &@ 980\two     & 10 &  1 &  1     & @--  &  2  & @\odlbyes  & round(36.762 ,1) & 2800+100+430     &                                   \\ \midrule
  4 &@ E5-2670v2       &@ 2 \mal 10 &@ --          &  8 &  5 &  1     & @--  &  1  & @\dlb      & round(21.382 ,1) & 4000-2*320       &                                   \\
    &@\ghz{2.5--3.3}   &            &@ 770         &  8 &  1 &  1     & @--  &  5  & @\nodlb    & round(35.9165,1) & 4000-320         &                                   \\
    &                  &            &@ 770\two     & 10 &  1 &  1     & @--  &  4  & @\dlb      & round(51.6925,1) & 4000             &                                   \\
  2 &                  &            &@ 780Ti       &  8 &  1 &  1     & @--  &  5  & @\odlbyes  & round(45.4965,1) & 4620-1*520       &                                   \\
    &                  &            &@ 780Ti\two   & 10 &  1 &  1     & @--  &  4  & @\dlb      & round(56.9215,1) & 4620             &                                   \\ %
    &                  &            &@ 780Ti\three & 10 &  1 &  1     & @--  &  4  & @\odlbyes  & round(61.0805,1) & 4620+1*520       &                                   \\
    &                  &            &@ 780Ti\four  & 10 &  1 &  1     & @--  &  4  & @\odlbyes  & round(64.436 ,1) & 4620+2*520       &                                   \\ \midrule
  2 &@ E5-2680v2       &@ 2 \mal 10 &@ --          &  8 &  2 &  2     &  8   &  1  & @\dlb      & round(26.798 ,1) & 4400             &                                   \\ %
    &@\ghz{2.8--3.6}   &            &@ K20X\two    &  8 &  1 &  1     & @--  &  5  & @\nodlb    & round(55.226 ,1) & 4400+2*2800      &                                   \\ %
    &                  &            &@ K40\two     &  8 &  1 &  1     & @--  &  5  & @\nodlb    & round(55.88  ,1) & 4400+2*3100      &                                   \\
    &                  &            &@ 980\fs      &  4 &  1 &  1     & @--  & 10  & @\dlb      & round(52.0455,1) & 4400+1*450       &                                   \\
    &                  &            &@ 980\fs\two  & 10 &  1 &  1     & @--  &  4  & @\dlb      & round(62.341, 1) & 4400+2*450       &                                   \\
    &                  &            &@ 980\fs\three& 10 &  1 &  1     & @--  &  4  & @\dlb      & round(65.0975,1) & 4400+3*450       &                                   \\
    &                  &            &@ 980\fs\four &  8 &  1 &  1     & @--  &  5  & @\nodlb    & round(66.9195,1) & 4400+4*450       &                                   \\ \midrule
  1 &@ AMD 6272 (2.1)  &@ 4 \mal 16 &@ --          &  5 &  5 &  2     & 14   &  1  & @\dlb      & round(23.6865,1) & 3670             &                                   \\ \midrule
  4 &@ AMD 6380        &@ 2 \mal 16 &@ --          &  5 &  5 &  1     &  7   &  1  & @\dlb      & round(16.031 ,1) & 2880             &                                   \\
    &@\ghz{2.5}        &            &@ TITAN       &  8 &  1 &  1     & @--  &  4  & @\nodlb    & round(32.5385,1) & 2880 + 750       &                                   \\
    &                  &            &@ 770\two     &  8 &  1 &  1     & @--  &  4  & @\dlb      & round(35.7535,1) & 2880 + 2*320     &                                   \\
    &                  &            &@ 980\fs      &  8 &  1 &  1     & @--  &  4  & @\nodlb    & round(35.6225,1) & 2880 + 1*450     &                                   \\
    &                  &            &@ 980\fs\two  &  8 &  1 &  1     & @--  &  4  & @\nodlb    & round(40.209 ,1) & 2880 + 2*450     &                                   \\
\bottomrule 
\end{spreadtab}
\end{center}
\end{table}

\begin{table}
\caption{Same as Table~\ref{tab:numbersAQP}, but for the \ac{RIB} benchmark.
\vspace{-7mm}}
\label{tab:numbersRIB}
\STautoround{2} %
\FPset\fRIB{3600}
\begin{center}
\myTableSettings
\begin{spreadtab}{{tabular}{c l c l r r r c r c P N{5}{0} S}} \toprule
@ U &@ processor(s)    &@CPUs \mal  &@ \acsp{GPU}  &@ \DDGrid         &@\npme &@\nth&@\mult{\ac{DLB}} &@\mult{$P$} &@\mult{{$\approx$\,cost}} & @\mult{\nsday per}   \\ 
@   &@\GHZ             &@cores      &@             &@x  &@y  &@z      &       &     &            &@\nanos     &@\euron           & @\mult{\fRIB\ \euro}              \\ \midrule
@ D &@ i7-4770K        &@ 1 \mal 4  &@  --         &  2 &  3 &  1     &  @--  &  1  & @\odlbyes  & 0.510      &  800             & \STcopy{v}{round([-2,0]/([-1,0]/\fRIB), 1)} \\ %
    &@\ghz{3.4--3.9}   &            &@ 980\FS      &  8 &  1 &  1     &  @--  &  1  & @\nodlb    & 1.3005     &  800+450         &                                   \\ \midrule
@ D &@ i7-5820K        &@ 1 \mal  6 &@  --         &  3 &  3 &  1     &   3   &  1  & @\dlb      & 0.691      &  850             &                                   \\ 
    &@\ghz{3.3--3.6}   &            &@ 770         &  4 &  1 &  1     &  @--  &  3  & @--        & 1.542      &  850+320         &                                   \\
    &                  &            &@ 980\FS      &  1 &  1 &  1     &  @--  & 12  & @--        & 1.799      &  850+540         &                                   \\ \midrule
  1 &@ E3-1270v2       &@ 1 \mal  4 &@  --         &  1 &  1 &  1     &  @--  &  8  & @--        & 0.301      & 1080             &                                   \\ %
    &@\ghz{3.5}        &            &@ 680         &  1 &  1 &  1     &  @--  &  8  & @--        & 0.886      & 1380             &                                   \\ 
    &                  &            &@ 770         &  1 &  1 &  1     &  @--  &  8  & @--        & 0.9065     & 1400             &                                   \\ \midrule %
  1 &@ E5-1620         &@ 1 \mal  4 &@ 680         &  1 &  1 &  1     &  @--  &  8  & @--        & 1.03       & 1900             &                                   \\ %
    &@\ghz{3.6--3.8}   &            &@ 770         &  1 &  1 &  1     &  @--  &  8  & @--        & 1.02       & 1900             &                                   \\ 
    &                  &            &@ 780         &  1 &  1 &  1     &  @--  &  8  & @--        & 1.06       & 1970             &                                   \\ %
    &                  &            &@ 780Ti       &  1 &  1 &  1     &  @--  &  8  & @--        & 1.14       & 1970+130         &                                   \\ %
    &                  &            &@ TITAN       &  1 &  1 &  1     &  @--  &  8  & @--        & 1.11       & 1900+430         &                                   \\ \midrule %
  1 &@ E5-1650         &@ 1 \mal  6 &@ 680         &  1 &  1 &  1     &  @--  & 12  & @--        & 1.09       & 2170             &                                   \\ %
    &@\ghz{3.2--3.8}   &            &@ 770         &  1 &  1 &  1     &  @--  & 12  & @--        & 1.13       & 2170             &                                   \\ %
    &                  &            &@ 780         &  1 &  1 &  1     &  @--  & 12  & @--        & 1.17       & 2170+70          &                                   \\ %
    &                  &            &@ 780Ti       &  1 &  1 &  1     &  @--  & 12  & @--        & 1.22       & 2370             &                                   \\ %
    &                  &            &@ 680\two     &  2 &  1 &  1     &  @--  &  6  & @\odlbyes  & 1.40       & 2470             &                                   \\ %
    &                  &            &@ 770\two     &  2 &  1 &  1     &  @--  &  6  & @\odlbyes  & 1.41       & 2470             &                                   \\ \midrule %
  1 &@ E5-2670         &@ 1 \mal  8 &@ 770         &  1 &  1 &  1     &  @--  & 16  & @--        & 1.39       & 2800             &                                   \\ %
    &@\ghz{2.6--3.3}   &            &@ 780         &  8 &  1 &  1     &  @--  &  2  & @\odlbyes  & 1.595      & 2800+70          &                                   \\ %
    &                  &            &@ 780Ti       &  1 &  1 &  1     &  @--  & 16  & @--        & 1.64       & 2800+200         &                                   \\ %
    &                  &            &@ TITAN       &  4 &  1 &  1     &  @--  &  4  & @\odlbyes  & 1.6685     & 2800+430         &                                   \\ %
    &                  &            &@ 770\two     &  2 &  1 &  1     &  @--  &  8  & @\odlbyes  & 1.72       & 3120             &                                   \\ \midrule %
  1 &@ E5-2670v2       &@ 1 \mal 10 &@ --          &  4 &  2 &  2     &   4   &  1  & @\dlb      & 0.788      & 2800-320         &                                   \\
    &@\ghz{2.5--3.3}   &            &@ 770         & 10 &  1 &  1     &  @--  &  2  & @\nodlb    & 1.7785     & 2800             &                                   \\
    &                  &            &@ 780         &  1 &  1 &  1     &  @--  & 20  & @--        & 1.60       & 2800+70          &                                   \\
    &                  &            &@ 780Ti       &  5 &  1 &  1     &  @--  &  4  & @\odlbyes  & 2.0635     & 2800+200         &                                   \\
    &                  &            &@ TITAN       &  1 &  1 &  1     &  @--  & 20  & @--        & 1.75       & 2800+430         &                                   \\ 
    &                  &            &@ 980         &  5 &  1 &  1     &  @--  &  4  & @\odlbyes  & 2.219      & 2800+100         &                                   \\
    &                  &            &@ 770\two     & 10 &  1 &  1     &  @--  &  2  & @\nodlb    & 2.1615     & 3120             &                                   \\
    &                  &            &@ 780Ti\two   &  4 &  1 &  1     &  @--  &  5  & @\odlbyes  & 2.3135     & 3120+2*200       &                                   \\
    &                  &            &@ 980\two     &  5 &  1 &  1     &  @--  &  4  & @\odlbyes  & 2.3365     & 2800+100+430     &                                   \\ \midrule
  4 &@ E5-2670v2       &@ 2 \mal 10 &@ --          &  8 &  2 &  2     &   8   &  1  & @\dlb      & 1.543      & 4000-2*320       &                                   \\
    &@\ghz{2.5--3.3}   &            &@ 770         & 20 &  1 &  1     &  @--  &  2  & @\nodlb    & 2.7115     & 4000-320         &                                   \\
    &                  &            &@ 770\two     &  8 &  5 &  1     &  @--  &  1  & @\nodlb    & 3.411      & 4000             &                                   \\
  2 &                  &            &@ 780Ti       &  8 &  5 &  1     &  @--  &  1  & @\odlbyes  & 3.3045     & 4620-1*520       &                                   \\
    &                  &            &@ 780Ti\two   &  8 &  1 &  1     &  @--  &  5  & @\nodlb    & 4.0215     & 4620             &                                   \\ %
    &                  &            &@ 780Ti\three &  8 &  5 &  1     &  @--  &  1  & @\odlbyes  & 4.174      & 4620+1*520       &                                   \\
    &                  &            &@ 780Ti\four  &  8 &  5 &  1     &  @--  &  1  & @\odlbyes  & 4.1725     & 4620+2*520       &                                   \\ \midrule
  2 &@ E5-2680v2       &@ 2 \mal 10 &@ --          & 10 &  3 &  1     &  10   &  1  & @\dlb      & 1.858      & 4400             &                                   \\ %
    &@\ghz{2.8--3.6}   &            &@ K20X\two    & 20 &  1 &  1     &  @--  &  2  & @\nodlb    & 3.986      & 4400+2*2800      &                                   \\ %
    &                  &            &@ K40\two     & 20 &  1 &  1     &  @--  &  2  & @\nodlb    & 4.0899     & 4400+2*3100      &                                   \\ %
    &                  &            &@ 980\fs      & 20 &  1 &  1     &  @--  &  2  & @\nodlb    & 3.991      & 4400+1*450       &                                   \\
    &                  &            &@ 980\fs\two  & 20 &  1 &  1     &  @--  &  2  & @\nodlb    & 4.688      & 4400+2*450       &                                   \\
    &                  &            &@ 980\fs\three& 20 &  1 &  1     &  @--  &  2  & @\nodlb    & 4.8495     & 4400+3*450       &                                   \\
    &                  &            &@ 980\fs\four & 20 &  1 &  1     &  @--  &  2  & @\nodlb    & 4.96       & 4400+4*450       &                                   \\ \midrule
  1 &@ AMD 6272 (2.1)  &@ 4 \mal 16 &@ --          &  5 &  5 &  2     &  14   &  1  & @\dlb      & 1.781      & 3670             &                                   \\ \midrule
  4 &@ AMD 6380        &@ 2 \mal 16 &@ --          &  5 &  5 &  1     &   7   &  1  & @\dlb      & 1.136      & 2880             &                                   \\
    &@\ghz{2.5}        &            &@ TITAN       & 16 &  2 &  1     &  @--  &  1  & @\nodlb    & 2.5785     & 2880 + 750       &                                   \\
    &                  &            &@ 770\two     & 16 &  1 &  1     &  @--  &  2  & @\nodlb    & 2.7395     & 2880 + 2*320     &                                   \\
    &                  &            &@ 980\fs      & 16 &  1 &  1     &  @--  &  2  & @\nodlb    & 2.812      & 2880 + 1*450     &                                   \\
    &                  &            &@ 980\fs\two  & 16 &  1 &  1     &  @--  &  2  & @\nodlb    & 3.1415     & 2880 + 2*450     &                                   \\
\bottomrule 
\end{spreadtab}
\end{center}
\end{table}

Tables~\ref{tab:numbersAQP} and \ref{tab:numbersRIB} list single-node performances 
for a diverse set of hardware combinations and the parameters that yielded peak
performance.
\QuotMarks{\acs{DD} grid} indicates the number of \ac{DD} cells per dimension,
whereas \QuotMarks{\nthread} gives the number of threads per rank.
Since each DD cell is assigned to exactly one MPI rank,
the total number of ranks can be calculated from the number of \ac{DD} grid
cells as \nrank $ = \text{DD}_x \times \text{DD}_y \times \text{DD}_z$ 
plus the number \npme of separate \ac{PME} ranks, if any. 
Normally the number of physical cores (or hardware threads with \ac{HT})
is the product of the number of ranks and the number of threads per rank.
For MPI parallel runs, the \ac{DLB} column indicates whether peak performance was
achieved with (symbol \dlb) or without \ac{DLB} (symbol \nodlb) or whether the benchmark was done exclusively 
with enabled \ac{DLB} (symbol \odlbyes).

The \QuotMarks{cost} column for each node gives a rough estimate on the net price as of 2014
and should be taken with a grain of salt.
Retail prices can easily vary by 15\,--\,20\percent over a relatively short period.
In order to provide a measure of \QuotMarks{bang for buck,}
using the collected cost and performance data we derive a performance-to-price ratio metric
shown in the last column. 
We normalize with the lowest performing setup to get $\ge 1$ values. 
While this ratio is only approximate, 
it still provides insight into which hardware combinations are significantly more
competitive than others. 

When a single CPU with 4\,--\,6 physical cores is combined with a single \ac{GPU},
using only threading without \ac{DD} resulted in the best performance.
On CPUs with 10 physical cores, peak performance was usually obtained with thread-MPI 
combined with multiple threads per rank.
When using multiple \acp{GPU},
where at least \nrank $ = N_\text{GPU}$ ranks is required,
in most cases an even larger number of ranks (multiple ranks per \ac{GPU}) was optimal.

\subsection{Speedup with \acp{GPU}}
Tables~\ref{tab:numbersAQP} and \ref{tab:numbersRIB} show that \acp{GPU} 
increase the performance of a compute node by a factor of 1.7\,--\,3.8.
In case of the inexpensive GeForce consumer cards, 
this also reflects in the node's performance-to-price ratio, 
which increases by a factor of 2\,--\,3 when adding at least one \ac{GPU} (last column).
When installing a significantly more expensive Tesla \ac{GPU}, 
the performance-to-price ratio is nearly unchanged.
Because both the performance itself (criterion C2, as defined in the introduction) 
as well as the performance-to-price ratio (C1) are so much better 
for nodes with consumer-class \acp{GPU}, 
we focused our efforts on nodes with this type of \ac{GPU}.

When looking at single-CPU nodes with one or more \acp{GPU}
(see third column of Tables~\ref{tab:numbersAQP} and \ref{tab:numbersRIB}),
the performance benefit obtained by a second \ac{GPU} is \unit[< 20]{\%}
for the \unit[80]{k} atom system (but largest on the 10-core machine), 
and on average about \unit[25]{\%} for the \unit[2]{M} atom system, 
whereas the performance-to-price ratio is nearly unchanged.

The dual-\ac{GPU}, dual-socket E5-2670v2 nodes are like the 
single-\ac{GPU}, single-socket E5-2670v2 nodes  
with the hardware of two nodes combined.
The dual-CPU nodes with several \acp{GPU} yielded the highest single-node performances of all tested nodes,
up to \unit[$\approx$ 67]{\nsday} for \ac{MEM} and \unit[$\approx$ 5]{\nsday} for \ac{RIB}
on the E5-2680v2 nodes with four GTX 980\fs.
The performance-to-price ratio (C1) of these 20-core nodes seems to have a sweet spot
at two installed \acp{GPU}.

\subsubsection*{\ac{GPU} application clock settings}
\label{pag:clocks}
\begin{figure}
\begin{center}
\includegraphics[width=15cm]{\pics 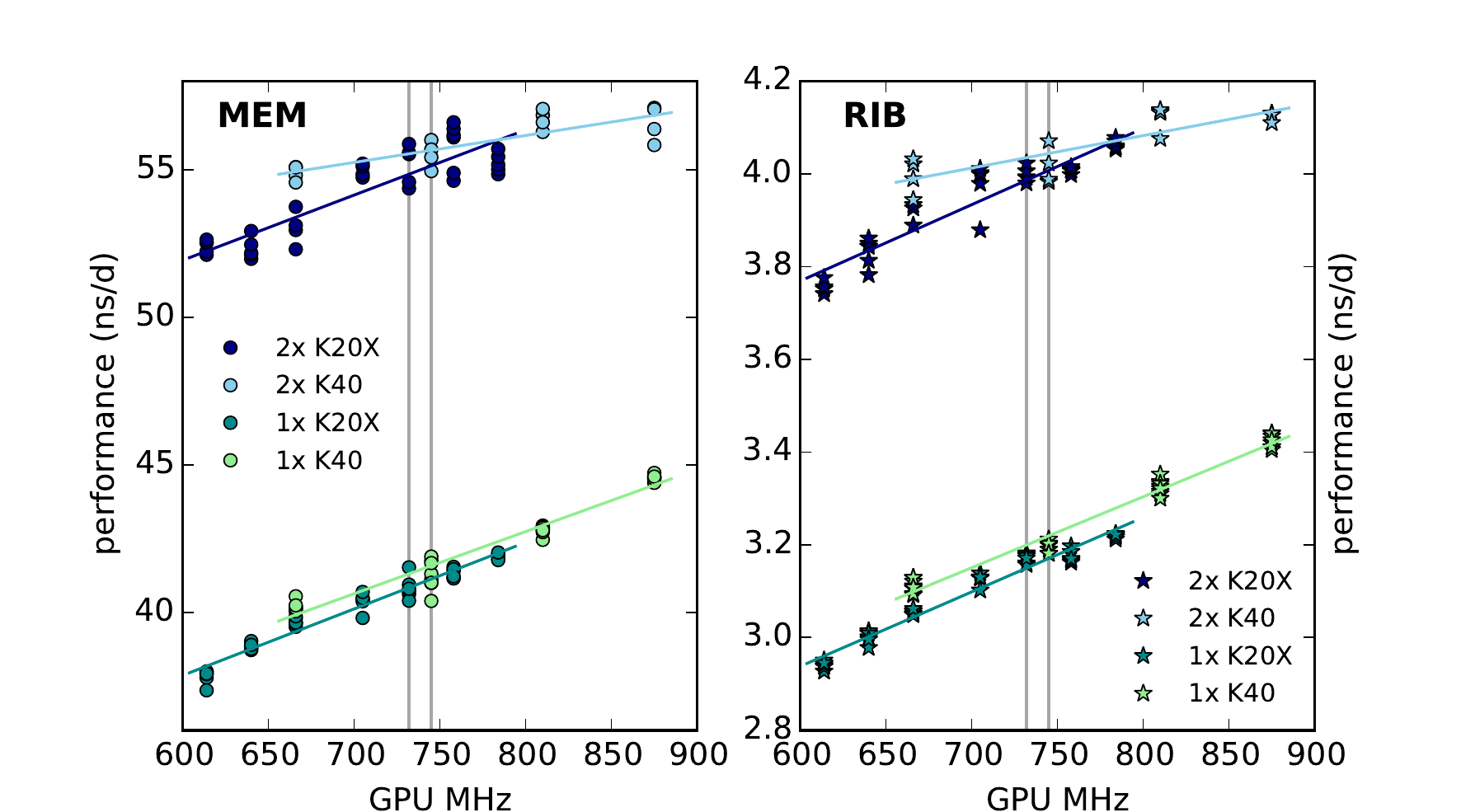}
\caption{
Performance as a function of the 
\ac{GPU} application clock rate on a node with 2\mal{}E5-2680v2 processors
and K20x (dark blue, dark green) or K40 (light blue, light green) \acp{GPU}.
Grey vertical lines indicate default clock rates.
\ac{MEM}, circles (\ac{RIB}, stars) benchmarks were run 
using the settings found in Table~\ref{tab:numbersAQP} (Table~\ref{tab:numbersRIB}).
}
\label{fig:clocks}
\end{center}
\end{figure}
For the Tesla K20X and K40 cards we determined how application clock settings 
influence the simulation performance
(as mentioned previously, GeForce cards do not support manual adjustment of the clock frequency).
While the default clock rate of the K20X is \unit[732]{MHz}, 
it supports seven clock rates in the range of \unit[614\,--\,784]{MHz}.
The K40 defaults to \unit[745]{MHz} and supports four rates in the range of \unit[666\,--\,875]{MHz}.
Application clock rates were set using NVIDIA's system management interface tool. 
\Eg, \type{nvidia-smi -ac 2600,875 -i 0}
sets the \ac{GPU} core clock rate to \unit[875]{MHz} 
and the memory clock rate to \unit[2600]{MHz} on interface 0. 

Fig.~\ref{fig:clocks} shows measured performances 
as a function of the clock rate  
as well as linear fits (lines).
For the K40, the maximum clock rate is about 17\percent higher 
than the default, and performance increases about 6.4\percent
when switching from the default to the maximum frequency using a single \ac{GPU}. 
The maximum clock rate of the K20X is about 7\percent higher than the default,
resulting in a 2.8\percent performance increase.
For two K40 or K20X \acp{GPU}, using the maximum clock rate only results in a 2.1\percent 
increased performance, likely because this hardware/benchmark combination is not \ac{GPU}-bound.
Highest performance is in all cases reached with the \ac{GPU} application 
clock set to the highest possible value.

\subsection{Energy efficiency}
\FPset\yrs{5}        %
\FPset\EpkWh{0.2}    %
\FPset\idleSevenW{27}
\FPset\idleNineW{24}
\FPset\idleNineWfs{24}

For a given CPU model, the performance/Watts ratio usually decreases with increasing clock
rate due to disproportionately higher energy dissipation.
CPUs with higher clock rates are therefore both more expensive
and less energy efficient. 
On the \ac{GPU} side, there has been a redesign for energy efficiency
with the Maxwell architecture,
providing significantly improved performance/Watt 
compared to Kepler generation cards.

\begin{table}[t b]
\begin{center}
\caption{Electric power consumption for nodes with up to 4 \acp{GPU}
when running the \ac{RIB} benchmark. 
Assuming \FPprint{\yrs} years of continuous operation and a price of \PreisEuro{\FPprint\EpkWh} per kWh including cooling,
the yield in produced trajectory per invested \PreisEuro{1000} is given in the last column.}
\label{tab:consumption}
\small
\STautoround{2}
\begin{spreadtab}{{tabular}{l l l N{5}{0} N{5}{0} N{5}{0} N{5}{0} N{5}{0}}} \toprule
@\mult{CPU}       &@\mult{installed}& \SThidecol &@\mult{RIB} &@\mult{production}                   &@ \mult{power }       &@ \mult{measured}   &@ \mult{power}                             &@ \mult{energy}                                  &@ \mult{node}          &@ \mult{trajectory}               &@ \mult{\yrs\ yr yield}                      \\ 
@\mult{cores}     &@  \acsp{GPU}    &            &@\nanos     &@\mult{($\mu${}s)}                   &@ \mult{draw (W)}     &@ \mult{(kWh/300s)} &@ \mult{draw (W)}                          &@ \mult{costs (\euro)}                           &@ \mult{costs (\euro)} &@ \mult{costs (\euro/$\mu$s)}     &@ \mult{(\text{ns}/k\euro)}                  \\ \midrule
@ E5-2670v2       &@  --            & 4          &@(node idle)&                                     &@ 135-140 \SThidecol  &  0.0115 \SThidecol & \STcopy{v}{g3*1000*12-(4-c3)*\idleSevenW} & \STcopy{v}{round(\yrs*h3*365*24*\EpkWh/1000, 0)}& 3360                  &@ \SThidecol                      &                                             \\
@ 2 \mal 10 c.    &@  --            & 0          &  1.38      & \STcopy{v}{round(d4*0.365*\yrs, 2)} &@ 378-386             &  0.031             &                                           &                                                 & 3360                  & \STcopy{v}{round((i4+j4)/e4, 0)} & \STcopy{v}{round(1000*e4/((i4+j4)/1000),0)} \\ \cmidrule(lr){2-7}
@ 2.5--3.3 GHz    &@ 780Ti          & 1          &  3.3045    &                                     &@ 548-633             &  0.051             &                                           &                                                 & 3880                  &                                  &                                             \\
@ (GCC 4.4.7)     &@ 780Ti  \mal 2  & 2          &  3.874     &                                     &@ 740-842             &  0.064             &                                           &                                                 & 4400                  &                                  &                                             \\
@                 &@ 780Ti  \mal 3  & 3          &  4.174     &                                     &@ 902-1018            &  0.080             &                                           &                                                 & 5430                  &                                  &                                             \\
@                 &@ 780Ti  \mal 4  & 4          &  4.1725    &                                     &@ 934-1045            &  0.078             &                                           &                                                 & 5950                  &                                  &                                             \\ \cmidrule(lr){2-7}
@                 &@ 980            & 1          &  3.678     &                                     &@ 514-549             &  0.044             & \STcopy{v}{g9*1000*12-(4-c9)*\idleNineW}  &                                                 & 3780                  &                                  &                                             \\
@                 &@ 980   \mal  2  & 2          &  4.182     &                                     &@ 614-696             &  0.053             &                                           &                                                 & 4200                  &                                  &                                             \\
@                 &@ 980   \mal  3  & 3          &  4.204     &                                     &@ 725-826             &  0.063             &                                           &                                                 & 5130                  &                                  &                                             \\
@                 &@ 980   \mal  4  & 4          &  4.1985    &                                     &@ 766-856             &  0.067             &                                           &                                                 & 5550                  &                                  &                                             \\ \midrule
@ E5-2680v2       &@  --            & 0          &@(node idle)&@                                    &@                     &@                   &  246 - 4 * \idleNineWfs                   &                                                 & 4400                  &@                                 &@                                            \\
@ 2 \mal 10 c.    &@  --            & 0          &  1.858     &                                     &@                     &                    &  542 - 4 * \idleNineWfs                   &                                                 & 4400                  &                                  &                                             \\ \cmidrule(lr){2-7}
@ 2.8--3.6 GHz    &@ 980\fs         & 1          &  3.991     &                                     &@                     &                    &  694 - 3 * \idleNineWfs                   &                                                 & 4400+1*450            &                                  &                                             \\
@ (GCC 4.8.3)     &@ 980\fs \mal 2  & 2          &  4.688     &                                     &@                     &                    &  847 - 2 * \idleNineWfs                   &                                                 & 4400+2*450            &                                  &                                             \\
@                 &@ 980\fs \mal 3  & 3          &  4.8495    &                                     &@                     &                    &  950 - 1 * \idleNineWfs                   &                                                 & 4400+3*450            &                                  &                                             \\
@                 &@ 980\fs \mal 4  & 4          &  4.96      &  \SThidecol                         &@                     &                    & 1092 - 0 * \idleNineWfs                   &                                                 & 4400+4*450            &                                  &                                             \\
\bottomrule
\end{spreadtab}
\end{center}
\end{table}
\begin{table}[t b]
\begin{center}
\caption{As Table~\ref{tab:consumption}, but for the \ac{MEM} benchmark}.
\label{tab:consumption2}
\small
\STautoround{2}
\begin{spreadtab}{{tabular}{l l l l N{5}{0} N{5}{0} c }} \toprule
@\mult{CPU}       &@\mult{installed} &@\mult{MEM} &@\mult{production}                   &@ \mult{power}           &@ \mult{energy}                                  &@ \mult{node}          &@ \mult{trajectory}               &@ \mult{\yrs\ yr yield}                  \\ 
@\mult{cores}     &@  \acsp{GPU}     &@\nanos     &@\mult{($\mu${}s)}                   &@ \mult{draw (W)}        &@ \mult{costs (\euro)}                           &@ \mult{costs (\euro)} &@ \mult{costs (\euro/$\mu$s)}     &@ \mult{($\mu$\text{s}/k\euro)}          \\ \midrule
@ E5-2680v2       &@ --              &  26.798    & \STcopy{v}{round(c3*0.365*\yrs, 2)} &  542 - 4 * \idleNineWfs & \STcopy{v}{round(\yrs*e3*365*24*\EpkWh/1000, 0)}& 4400                  & \STcopy{v}{round((f3+g3)/d3, 0)} & \STcopy{v}{round(d3/((f3+g3)/1000), 3)} \\ \cmidrule(lr){2-7}
@ 2 \mal 10 c.    &@ 980\fs          &  52.0455   &                                     &  619 - 3 * \idleNineWfs &                                                 & 4400+1*450            &                                  &                                         \\
@ 2.8--3.6 GHz    &@ 980\fs \mal 2   &  62.341    &                                     &  773 - 2 * \idleNineWfs &                                                 & 4400+2*450            &                                  &                                         \\
@ (GCC 4.8.3)     &@ 980\fs \mal 3   &  65.0975   &                                     &  848 - 1 * \idleNineWfs &                                                 & 4400+3*450            &                                  &                                         \\
@                 &@ 980\fs \mal 4   &  66.9195   & \SThidecol                          &  899 - 0 * \idleNineWfs &                                                 & 4400+4*450            &  \SThidecol                      &                                         \\
\bottomrule
\end{spreadtab}
\end{center}
\end{table}
For several nodes with decent performance-to-price ratios
from Tables~\ref{tab:numbersAQP}\,--\,\ref{tab:numbersRIB} 
we determined the energy efficiency 
by measuring the power consumption when running the
benchmarks at optimal settings (Tables~\ref{tab:consumption} and \ref{tab:consumption2}).
On the E5-2670v2 nodes, we measured the average power draw 
over an interval of 300 seconds
using a Voltcraft \QuotMarks{EnergyCheck 3000} meter. 
On the E5-2680v2 nodes, the current energy consumption was read from the power supply
with \type{ipmitool}.\footnote{\url{http://ipmitool.sourceforge.net/}} 
We averaged over 100 readouts one second apart each.
Power measurements were taken after the initial load balancing phase.
For the power consumption of idle \acp{GPU},
we compared the power draw of idle nodes with and without four installed cards,
resulting in \unit[$\approx$\,27]{W} (\unit[$\approx$\,24]{W}) for a single idle 780Ti (980).

While the total power consumption of nodes without \acp{GPU} is lowest,
their trajectory costs are the highest due to the very low trajectory production rate.
Nodes with one or two \acp{GPU} produce about 1.5\,--\,2\mal as much \ac{MD} trajectory
per invested \euro\ than CPU-only nodes 
(see last column in Tables~\ref{tab:consumption} and \ref{tab:consumption2}).
While trajectory production is cheapest with
one or two \acp{GPU}, due to the runs becoming CPU-bound, 
the cost rises significantly with the third or fourth card, 
though it does not reach the CPU-only level.
To measure the effect of \ac{GPU} architectural change on the energy efficiency of a node,
the E5-2670v2 node was tested both with GTX 780Ti (Kepler) and GTX 980 (Maxwell) cards.
When equipped with 1\,--\,3 \acp{GPU}, 
the node draws \unit[$>$100]{W} less power under load 
using Maxwell generation cards than with Kepler.
This results in about 20\% reduction of trajectory costs, lowest for the node with
two E5-2670v2 CPUs combined with a single GTX 980 \ac{GPU}.
Exchanging the E5-2670v2 with E5-2680v2 CPUs, 
which have $\approx$10\percent higher clock frequency, 
yields a 52\percent (44\percent) increase in energy consumption and 30\percent (21\percent) higher
trajectory costs for the case of one \ac{GPU} (two \acp{GPU}).

\subsubsection{Well-balanced CPU\,/\,GPU resources are crucial}
With \acp{GPU}, the short-range pair interactions are
offloaded to the \ac{GPU}, while the calculation of other interactions like
bonded forces, constraints, and the \ac{PME} mesh, 
remains on the CPU.
To put all available compute power to optimum use,
\gromacs balances the load between CPU and \ac{GPU}
by shifting as much computational work as possible from the \ac{PME} mesh part to the short-range electrostatic kernels.
As a consequence of the offload approach, the achievable performance is limited by the time 
spent by the CPU in the non-overlapping computation where the \ac{GPU} is left idle, like 
constraints calculation, integration, neighbor search and domain-decomposition.

\begin{table}[t b]
\begin{center}
\caption{Dependence of simulation performance $P$, cutoff settings, 
and total power consumption on the graphics processing power
for the \ac{RIB} system on a node with 2{}\mal 2680v2 CPUs and up to 4 GTX 980\fs\ \acp{GPU}.
The \QuotMarks{cost ratios} indicate the floating point operations in this part of
the calculation relative to the CPU-only case.}
\label{tab:example}
\STautoround{2}
\FPset\SrMflopsPerStep{55920}
\FPset\PmeMflopsPerStep{6581}
\begin{spreadtab}{{tabular}{c l N{5}{0} l l l c}} \toprule
@\mult{installed}&@\mult{$P$}       &@\mult{tot. power} &@\mult{cutoff} &@\mult{cost ratio}  &@\mult{cost ratio}         &@\mult{cost ratio}        &@\mult{energy efficiency}    \\ 
@  \acsp{GPU}    &@\nanos           &@\mult{draw (W)}   &@\mult{(nm)}   &@\mult{short range} &@\mult{short range}        &@\mult{PME 3D FFT}        &@\mult{(W/ns/d)}             \\ \midrule
@ 0              & round(1.858 , 2) &  446              & 1.0           &  1.0 \SThidecol    &  55920 / \SrMflopsPerStep & 6581 / \PmeMflopsPerStep & \STcopy{v}{round(c3/b3, 0)} \\
@ 1              & round(3.991 , 2) &  622              & 1.157         &  1.54              & 145633 / \SrMflopsPerStep & 4267 / \PmeMflopsPerStep &                             \\
@ 2              & round(4.688 , 2) &  799              & 1.378         &  2.59              & 209766 / \SrMflopsPerStep & 3012 / \PmeMflopsPerStep &                             \\
@ 3              & round(4.8495, 2) &  926              & 1.447         &  2.99              & 236115 / \SrMflopsPerStep & 2779 / \PmeMflopsPerStep &                             \\
@ 4              & round(4.96  , 2) & 1092              & 1.607         &  4.1               & 299915 / \SrMflopsPerStep & 2356 / \PmeMflopsPerStep &                             \\ \midrule
\end{spreadtab}
\end{center}
\end{table}

Table~\ref{tab:example} shows 
the distribution of the \ac{PME} and short-range non-bonded workload with increasing graphics processing power.
Adding the first \ac{GPU} relieves the CPU from the complete 
short-range non-bonded calculation.
Additionally, load balancing shifts work from the \ac{PME} mesh (CPU) to the
non-bonded kernels (\ac{GPU}), so that the CPU spends less time
in the \ac{PME} 3D FFT calculation. Both effects yield a 2.1\mal higher performance
compared to the case without a \ac{GPU}.
The benefit of additional \acp{GPU} 
is merely the further reduction of the \ac{PME} 3D FFT workload (which is just part
of the CPU workload) by a few extra percent. 
The third and fourth \ac{GPU} only reduce the CPU workload by a tiny amount,
resulting in a few percent extra performance.
At the same time, the \ac{GPU} workload is steadily increased,
and with it increases the \ac{GPU} power draw,
reflecting in a significantly increased power consumption of the node
(see also the other 3- and 4-\ac{GPU} benchmarks in Table~\ref{tab:consumption}).

Hence, \ac{GPU} and CPU resources should always be chosen in tandem
keeping in mind the needs of the intended simulation setup.
More or faster \acp{GPU} will have little effect
when the bottleneck is on the CPU side.
The theoretical peak throughput as listed in Table~\ref{tab:NVIDIA}
helps to roughly relate different \ac{GPU} configurations to each other
in terms of how much \ac{SP} compute power they provide 
and of how much CPU compute power is needed to achieve a balanced hardware setup.
At the lower end of \ac{GPU} models studied here are 
the Kepler GK104 cards: GTX 680 and 770.
These are followed by GK110 cards, in 
order of increasing compute power, 
GTX 780, K20X, GTX TITAN, K40, GTX 780Ti.
The GTX 980 and TITAN X, based on the recent Maxwell architecture, are the fastest 
as well as most power-efficient GPUs tested.
The dual-chip server-only Tesla K80 can provide even higher performance on a single board.

The performance-to-price ratios presented in this study reflect
the characteristics of the specific combinations of 
hardware setups and workloads used in our benchmarks.
Different types of simulations or input setups may expose slightly different
ratios of CPU to \ac{GPU} workload. 
\Eg, when comparing a setup using the AMBER force-field with \unit[0.9]{nm} cutoffs
to a CHARMM setup with \unit[1.2]{nm} cutoffs and switched van der Waals interactions,
the latter results in a larger amount of pair interactions to be computed, hence more \ac{GPU} workload.
This in turn leads to slight differences in the ideal CPU-\ac{GPU} balance in these two cases.
Still, given the representative choices of hardware configurations and simulation systems, our results allow for drawing general conclusions about similar simulation setups.

\subsection{Multi-simulation throughput}
Our general approach in this study is using single-simulation benchmarks for hardware evaluation. 
However, comparing the performance $P$ on a node with just a few cores
to $P$ on a node with many cores (and possibly several \acp{GPU})
is essentially a strong scaling scenario involving efficiency reduction due to
MPI communication overhead and/or lower multi-threading efficiency. 

This is alleviated by partitioning available processor cores between multiple 
replicas of the simulated system, which \eg differ in their starting configuration.
Such an approach is generally useful if average properties of a simulation ensemble
are of interest. With several replicas, the parallel efficiency is higher,
since each replica is distributed to fewer cores.
A second benefit is a higher \ac{GPU} utilization due to \ac{GPU} sharing. 
As the individual replicas do not run completely synchronized,
the fraction of the time step that the \ac{GPU} is normally left idle 
is used by other replicas. The third benefit, similar to the case of 
\ac{GPU} sharing by ranks of a single simulation, is that independent simulations 
benefit from \ac{GPU} task overlap if used in conjunction with CUDA MPS.
In effect, CPU and \ac{GPU} resources are both used more efficiently,
at the expense of getting multiple shorter trajectories instead of a single long one.

\begin{figure}[tb]
\begin{center}
\includegraphics[width=\textwidth]{\pics 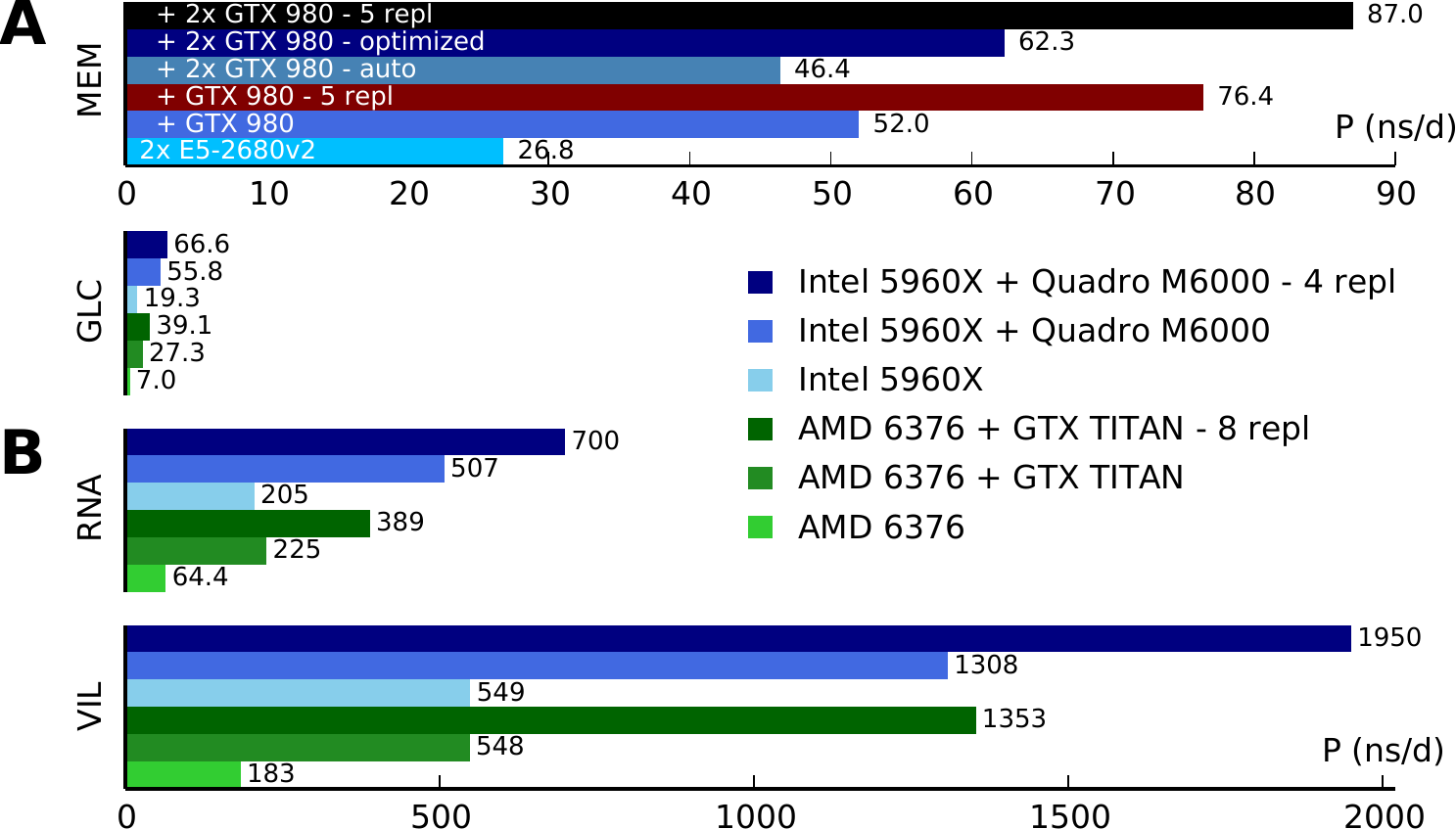}
\caption{Maximizing throughput by running multiple simulations per node.
\textbf{A}: Single-simulation performance $P$ of the \ac{MEM} benchmark on a node with 2\mal{}E5-2680v2 CPUs using
0, 1, or 2 GTX 980\fs \acp{GPU} (blue colors) compared to the aggregated performance
of 5 replicas (red\,/\,black).
\textbf{B}: Similar to A, but for different node types and benchmark systems
(available at 
\type{http://www.gromacs.org/gpu} and 
\type{ftp://ftp.gromacs.org/pub/CRESTA/CRESTA\_Gromacs\_benchmarks\_v2.tgz}).
GLC -- \unit[144]{k} atoms GluCL CRESTA benchmark, \unit[1]{nm} cutoffs, \ac{PME} grid spacing \unit[0.12]{nm}.
RNA -- \unit[14.7]{k} atoms solvated RNAse, \unit[0.9]{nm} cutoffs, \ac{PME} grid spacing \unit[0.1125]{nm}.
VIL -- \unit[8]{k} atoms villin protein, \unit[1]{nm} cutoffs, \ac{PME} grid spacing \unit[0.125]{nm}.
In B a \unit[5]{fs} time step and \gromacs 5.0.4 was used.
}
\label{fig:throughput}
\end{center}
\end{figure}
Figure~\ref{fig:throughput} quantifies this effect for small to medium \ac{MD} systems. %
Subplot A compares the the \ac{MEM} performance 
for a single-simulation (blue colors)
to the aggregated performance of five replicas (red\,/\,black).
The aggregated trajectory production of a multi-simulation 
is the sum of the produced trajectory lengths of the individual replicas.
The single simulations settings are found in Table~\ref{tab:numbersAQP};
in multi-simulations we used one rank with
40 / \nrank threads per replica.  
For a single GTX 980, the aggregated performance of a 5-replica simulation (red bar)
is 47\percent higher than the single simulation optimum.
While there is a performance benefit of $\approx$25\percent already for two replicas,
the effect is more pronounced for $\ge4$ replicas.
For two 980 \acp{GPU}, the aggregated performance of five replicas is 
40\percent higher than the performance of a single simulation at optimal settings
or 87\percent higher when compared to a single simulation at default settings
(\nrank = 2, \nthread = 20).

Subplot B compares single and multi-simulation throughput for \ac{MD} systems 
of different size for an octacore Intel (blue bars) %
and a 16-core AMD node (green bars). %
Here, within each replica we used OpenMP threading exclusively,
with the total number of threads being equal to the number of cores of the node.
The benefit of multi-simulations is always significant 
and more pronounced the smaller the \ac{MD} system. 
It is also more pronounced on the AMD Opteron processor as compared to the Core i7 architecture.
For the \unit[8]{k} atom VIL example, the performance gain is nearly a factor of 2.5
on the 16-core AMD node. 

As with multi-simulations one essentially shifts resource use from strong scaling
to the embarrassingly parallel scaling regime, 
the benefits increase the smaller the input system, the larger the number of CPU cores per \ac{GPU},
and the worse the single-simulation CPU-\ac{GPU} overlap.

Section~\ref{sec:multi} in the supporting information 
gives examples of multi-simulation setups in \gromacs
and additionally quantifies the performance benefits of multi-simulations
across many nodes connected by a fast network.

\subsection{Strong scaling}
The performance $P$ across multiple nodes is given in Tables~\ref{tab:scalingAQP}\,--\,\ref{tab:scalingRIB} 
for selected hardware configurations.
The parallel efficiency $E$ is the performance on $m$ nodes 
divided by $m$ times the performance on a single node: 
$E_m = P_m / (m \times P_1)$.
In the spirit of pinpointing the highest possible performance for each hardware combination,
the multi-node benchmarks were done with a standard MPI library,
whereas on individual nodes the low-overhead and therefore faster thread-MPI implementation was used.
This results in a more pronounced drop in parallel efficiency from a single to  many nodes
than what would be observed when using a standard MPI library throughout.
The prices in Tables~\ref{tab:numbersAQP}\,--\,\ref{tab:numbersRIB} 
do neither include an \acf{IB} network adapter
nor proportionate costs for an \ac{IB} switch port. 
Therefore, the performance-to-price ratios are slightly lower 
for nodes equipped for parallel operation as compared to the values in the tables.
However, the most important factor limiting the performance-to-price ratio for parallel operation
is the parallel efficiency that is actually achieved. 
 
The raw performance of the \ac{MEM} system can exceed \unit[300]{\nsday} on state-of-the-art hardware,
and also the bigger \ac{RIB} system exceeds \unit[200]{\nsday}.
This minimizes the time-to-solution, however at the expense of the parallel efficiency $E$
(last column).
Using activated \ac{DLB} and separate \ac{PME} ranks
yielded the best performance on CPU-only nodes throughout.
With \acp{GPU} the picture is a bit more complex. 
On large node counts, a homogeneous, interleaved \ac{PME} rank distribution
showed a significantly higher performance than without separate \ac{PME} ranks. 
\ac{DLB} was beneficial only for the \ac{MEM} system on small numbers of \ac{GPU} nodes.
\ac{HT} helped also across several nodes in the low- to medium-scale regime,
but not when approaching the scaling limit. The performance benefits from
\ac{HT} are largest on individual nodes and in the range of 5\,--\,15\%.

\begin{table}
\begin{minipage}{\textwidth}
\renewcommand{\footnoterule}{} %
\renewcommand{\thefootnote}{\alph{footnote}}
\newcommand{\fnm}{\footnotemark[1]}
\caption{Scaling of the \ac{MEM} benchmark on different node types with
performance $P$ and parallel efficiency $E$.
A black \QuotMarks{\hy} symbol indicates that using all hyper-threading cores resulted in the fastest execution, 
otherwise using only the physical core count was more advantageous. 
A grey \QuotMarks{\HY} denotes that this benchmark was done only with the hyper-threading core count (= 2\mal physical).
}
\label{tab:scalingAQP}

\newcommand{\sss}{\hspace{3ex}}
\STautoround{2}
\begin{center}
\small
\begin{spreadtab}{{tabular}{r l l r r r c c c c P l l}} \toprule
@No. of&@ processor(s)          &@\acsp{GPU},&@ \DDGrid     &@  \npme\,/  &@\nth&     &@\mult{DLB} & @\mult{$P$}        & @\mult{$E$}                 \\
@nodes &@ Intel                 &@Infiniband &@x  &@y  &@z  &@  node      &     &     &            & @\mult{(\nsday)}   &                             \\ \midrule
   1   &@ E3-1270v2             &@ 770,      &  1 &  1 &  1 &   @--       &   8 &@\HY & @--        & round(  20.530, 1) & \STcopy{v}{k3/(!a3*k!3)}    \\
   2   &@\sss (4 cores)         &@ QDR\fnm   &  2 &  1 &  1 &   @--       &   8 &@\HY & @\odlbyes  & round(  27.218, 1) &                             \\
   4   &                        &            &  4 &  1 &  1 &   @--       &   8 &@\HY & @\odlbyes  & round(  22.077, 1) &                             \\
   8   &                        &            &  8 &  1 &  1 &   @--       &   8 &@\HY & @\odlbyes  & round(  68.311, 1) &                             \\
  16   &                        &            & 16 &  1 &  1 &   @--       &   8 &@\HY & @\odlbyes  & round(  85.705, 1) &                             \\
  32   &                        &            &  8 &  4 &  1 &   @--       &   8 &@\HY & @\odlbyes  & round( 119.229, 0) &                             \\ \midrule
   1   &@ E5-1620               &@ 680,      &  1 &  1 &  1 &   @--       &   8 &@\hy & @--        & round( 21.0375, 1) & \STcopy{v}{k9/(!a9*k!9)}    \\
   2   &@\sss (4 cores)         &@ QDR       &  2 &  1 &  1 &   @--       &   8 &@\hy & @\odlbyes  & round(  28.950, 1) &                             \\
   4   &                        &            &  4 &  1 &  1 &   @--       &   8 &@\hy & @\odlbyes  & round(  46.925, 1) &                             \\ \midrule
   1   &@ E5-2670v2             &@ 780Ti\two,& 10 &  1 &  1 &   @--       &   4 &@\hy & @\dlb      & round(  56.9215,1) & \STcopy{v}{k12/(!a12*k!12)} \\          %
   2   &@\sss (2\mal{}10 cores) &@ QDR       &  4 &  5 &  1 &   @--       &   2 &@    & @\dlb      & round(  74.2125,1) &                             \\          %
   4   &                        &            &  8 &  1 &  1 &   8/[-6,0]  &   5 &@    & @\nodlb    & round( 103.379, 1) &                             \\          %
   8   &                        &            &  8 &  1 &  2 &  16/[-6,0]  &   5 &@    & @\nodlb    & round( 119.1125,1) &                             \\          %
  16   &                        &            &  8 &  4 &  1 &  32/[-6,0]  &   5 &@    & @\nodlb    & round( 164.767 ,1) &                             \\          %
  32   &                        &            &  8 &  8 &  1 &  64/[-6,0]  &   5 &@    & @\nodlb    & round( 193.123, 1) &                             \\ \midrule %
   1   &@  E5-2670v2            &@ 980\two,  & 10 &  1 &  1 &   @--       &   4 &@\hy & @\odlbyes  & round(  58.035 ,1) & \STcopy{v}{k18/(!a18*k!18)} \\
   2   &@\sss (2\mal{}10 cores) &@ QDR       &  4 &  5 &  1 &   @--       &   2 &@    & @\odlbyes  & round(  75.5715,1) &                             \\
   4   &                        &            &  8 &  5 &  1 &   @--       &   2 &@    & @\nodlb    & round(  96.624, 1) &                             \\ \midrule
   1   &@ E5-2680v2             &@ --        &  8 &  2 &  2 &   8/[-6,0]  &   1 &@\hy & @\dlb      & round(  26.798, 1) & \STcopy{v}{k21/(!a21*k!21)} \\ 
   2   &@\sss (2\mal{}10 cores) &@ FDR-14,   &  4 &  5 &  3 &  20/[-6,0]  &   1 &@\hy & @\dlb      & round(  42.000, 1) &                             \\
   4   &                        &            &  8 &  5 &  3 &  40/[-6,0]  &   1 &@\hy & @\dlb      & round(  76.348, 1) &                             \\
   8   &                        &            &  8 &  7 &  2 &  48/[-6,0]  &   2 &@\hy & @\dlb      & round( 122.379, 0) &                             \\
   16  &                        &            &  8 &  8 &  4 &  64/[-6,0]  &   1 &@    & @\dlb      & round( 161.857, 0) &                             \\
   32  &                        &            &  8 &  8 &  8 & 128/[-6,0]  &   1 &@    & @\dlb      & round( 209.235, 0) &                             \\
   64  &                        &            & 10 &  8 &  6 & 160/[-6,0]  &   2 &@    & @\dlb      & round( 240.175, 0) &                             \\ \midrule
   1   &@ E5-2680v2             &@ K20X\two  &  8 &  1 &  1 &   @--       &   5 &@\hy & @\dlb      & round(  55.226, 1) & \STcopy{v}{k28/(!a28*k!28)} \\
   2   &@\sss (2\mal{}10 cores) &@ (732 MHz),&  4 &  5 &  1 &   @--       &   4 &@\hy & @\dlb      & round(  74.544, 1) &                             \\
   4   &                        &@ FDR-14    &  8 &  1 &  2 &   @--       &   5 &@    & @\dlb      & round( 118.348, 0) &                             \\
   8   &                        &            &  8 &  1 &  2 &  16/[-6,0]  &   5 &@    & @\nodlb    & round( 162.528, 0) &                             \\
  16   &                        &            &  8 &  4 &  1 &  32/[-6,0]  &   5 &@    & @\nodlb    & round( 226.001, 0) &                             \\
  32   &                        &            &  8 &  8 &  1 &  64/[-6,0]  &   5 &@    & @\nodlb    & round( 303.926, 0) &                             \\
\bottomrule
\end{spreadtab}
\end{center}
\footnotetext[1]{Note: \pcixcomment}
\end{minipage}
\end{table}

\begin{table}
\begin{minipage}{\textwidth}
\renewcommand{\footnoterule}{} %
\renewcommand{\thefootnote}{\alph{footnote}}
\newcommand{\fnm}{\footnotemark[1]}
\caption{Same as Table~\ref{tab:scalingAQP}, but for the \ac{RIB} benchmark.}
\label{tab:scalingRIB}
\newcommand{\sss}{\hspace{3ex}}
\STautoround{2}
\begin{center}
\small
\begin{spreadtab}{{tabular}{r l l r r r c r c c P l l}} \toprule
@No. of&@ processor(s)          &@\acsp{GPU},&@ \DDGrid     &@ \npme\,/   &@\nth&     &@\mult{\ac{DLB}} & @\mult{$P$}  & @\mult{$E$}                 \\
@nodes &@ Intel                 &@Infiniband &@x  &@y  &@z  &@ node       &     &     &            & @\mult{(\nsday)}  &                             \\ \midrule
   1   &@ E3-1270v2             &@ 770,      &  1 &  1 &  1 &   @--       &  8  &@\HY & @--        & round(  0.907, 2) & \STcopy{v}{k3/(!a3*k!3)}    \\
   2   &@\sss (4 cores)         &@ QDR\fnm   &  2 &  1 &  1 &   @--       &  8  &@\HY & @\odlbyes  & round(  1.866, 2) &                             \\
   4   &                        &            &  4 &  1 &  1 &   @--       &  8  &@\HY & @\odlbyes  & round(  2.991, 2) &                             \\
   8   &                        &            &  8 &  1 &  1 &   @--       &  8  &@\HY & @\odlbyes  & round(  4.929, 2) &                             \\
  16   &                        &            & 16 &  1 &  1 &   @--       &  8  &@\HY & @\odlbyes  & round(  4.741, 2) &                             \\
  32   &                        &            & 16 &  2 &  1 &   @--       &  8  &@\HY & @\odlbyes  & round( 10.251, 1) &                             \\ \midrule
   1   &@ E5-2670v2             &@ 780Ti\two,&  8 &  1 &  1 &   @--       &  5  &@\HY & @\nodlb    & round(  4.0215,2) & \STcopy{v}{k9/(!a9*k!9)}    \\          %
   2   &@\sss (2\mal{}10 cores) &@ QDR       & 20 &  1 &  1 &   @--       &  4  &@\hy & @\nodlb    & round(  6.225, 2) &                             \\          %
   4   &                        &            &  8 &  5 &  1 &   @--       &  4  &@\hy & @\nodlb    & round( 10.7615,2) &                             \\          %
   8   &                        &            & 16 & 10 &  1 &   @--       &  2  &@\hy & @\nodlb    & round( 16.554, 2) &                             \\          %
  16   &                        &            & 16 & 10 &  1 &   @--       &  2  &     & @\nodlb    & round( 23.775, 2) &                             \\          %
  32   &                        &            & 16 & 10 &  2 &   @--       &  2  &     & @\nodlb    & round( 33.512, 2) &                             \\ \midrule %
   1   &@  E5-2670v2            &@ 980\two,  &  8 &  5 &  1 &   @--       &  1  &@\hy & @\odlbyes  & round(  4.182, 2) & \STcopy{v}{k15/(!a15*k!15)} \\
   2   &@\sss (2\mal{}10 cores) &@ QDR       & 20 &  1 &  1 &   @--       &  4  &@\hy & @\nodlb    & round( 6.6045, 2) &                             \\ 
   4   &                        &            &  8 &  5 &  1 &   @--       &  4  &@\hy & @\nodlb    & round(10.9565, 1) &                             \\ \midrule
   1   &@ E5-2680v2             &@ --        & 10 &  3 &  1 &   10/[-6,0] &  1  &@\hy & @\dlb      & round(  1.858, 2) & \STcopy{v}{k18/(!a18*k!18)} \\ 
   2   &@\sss (2\mal{}10 cores) &@ FDR-14    & 10 &  3 &  1 &   10/[-6,0] &  2  &@\hy & @\dlb      & round(  3.239, 2) &                             \\
   4   &                        &            & 10 &  2 &  3 &   20/[-6,0] &  2  &@\hy & @\dlb      & round(  6.123, 2) &                             \\
   8   &                        &            &  8 &  5 &  3 &   40/[-6,0] &  2  &@\hy & @\dlb      & round( 12.253, 1) &                             \\
   16  &                        &            & 10 &  8 &  3 &   80/[-6,0] &  2  &@\hy & @\dlb      & round( 21.776, 1) &                             \\
   32  &                        &            & 10 &  7 &  7 &  150/[-6,0] &  2  &@\hy & @\dlb      & round( 39.425, 1) &                             \\
   64  &                        &            & 16 & 10 &  6 &  320/[-6,0] &  1  &     & @\dlb      & round( 70.743, 1) &                             \\
   128 &                        &            & 16 & 16 &  8 &  512/[-6,0] &  1  &     & @\dlb      & round(127.66 , 0) &                             \\
   256 &                        &            & 16 & 17 & 15 & 1040/[-6,0] &  1  &     & @\dlb      & round(185.950, 0) &                             \\
   512 &                        &            & 20 & 16 & 13 &  960/[-6,0] &  2  &     & @\nodlb    & round(207.543, 0) &                             \\ \midrule
   1   &@ E5-2680v2             &@ K20X\two  & 20 &  1 &  1 &  @--        &  2  &@\hy & @\nodlb    & round(  3.986, 2) & \STcopy{v}{k28/(!a28*k!28)} \\
   2   &@\sss (2\mal{}10 cores) &@ (732 MHz),& 10 &  8 &  1 &  @--        &  1  &@\hy & @\nodlb    & round(  5.014, 2) &                             \\
   4   &                        &@ FDR-14    & 10 &  8 &  1 &  @--        &  2  &@\hy & @\nodlb    & round(  9.526, 2) &                             \\
   8   &                        &            & 16 & 10 &  1 &  @--        &  2  &@\hy & @\nodlb    & round( 16.247, 1) &                             \\
  16   &                        &            & 16 & 10 &  1 &  @--        &  2  &     & @\nodlb    & round( 27.475, 1) &                             \\
  32   &                        &            &  8 &  8 &  1 &   64/[-6,0] &  5  &     & @\nodlb    & round( 49.095, 1) &                             \\
  64   &                        &            & 16 &  8 &  1 &  128/[-6,0] &  5  &     & @\nodlb    & round( 85.332, 1) &                             \\
 128   &                        &            & 16 & 16 &  1 &  256/[-6,0] &  5  &     & @\nodlb    & round(129.667, 1) &                             \\
 256   &                        &            & 16 &  8 &  4 &  512/[-6,0] &  5  &     & @\nodlb    & round(139.535, 1) &                             \\
\bottomrule
\end{spreadtab}
\end{center}
\footnotetext[1]{Note: \pcixcomment}
\end{minipage}
\end{table}

The E3-1270v2 nodes with QDR \ac{IB} exhibit an unexpected, erratic scaling behaviour
(see Tables~\ref{tab:scalingAQP}\,--\,\ref{tab:scalingRIB}, top rows).
The parallel efficiency is not decreasing strictly monotonic, as one would expect.
\label{pag:intelE3nodes}
The reason could be the CPU's limited number of 20 \ac{PCIe} lanes,
of which 16 are used by the \ac{GPU}, leaving
only 4 for the \ac{IB} adapter. However, the QDR \ac{IB} adapter requires
8 \ac{PCIe} 2.0 lanes to exploit the full QDR bandwidth.
This was also verified in an MPI bandwidth test between two of these nodes (not shown).
Thus, while the E3-1270v2 nodes with \ac{GPU} offer an attractive performance-to-price ratio,
they are not well suited for parallel operation.
Intel's follow-up model, the E3-1270v3
provides only 16 \ac{PCIe} lanes, just enough for a single \ac{GPU}.
For parallel usage, the processor models of the E5-16x0, E5-26x0 and E5-26x0v2
are better suited as they offer 40 \ac{PCIe} lanes, 
enough for two \acp{GPU} plus \ac{IB} adapter.

\section{Discussion}
A consequence of offloading the short-ranged non-bonded forces to graphics card(s)
is that performance depends on the ratio between CPU and \ac{GPU} compute power.
This ratio can therefore be optimized,
depending on the requirements of the simulation systems.
Respecting that, for any given CPU configuration there is an optimal
amount of \ac{GPU} compute power for most economic trajectory production,
which depends on energy and hardware costs.

\begin{figure}[t b p]
\begin{center}
\includegraphics[width=16.5cm]{\pics 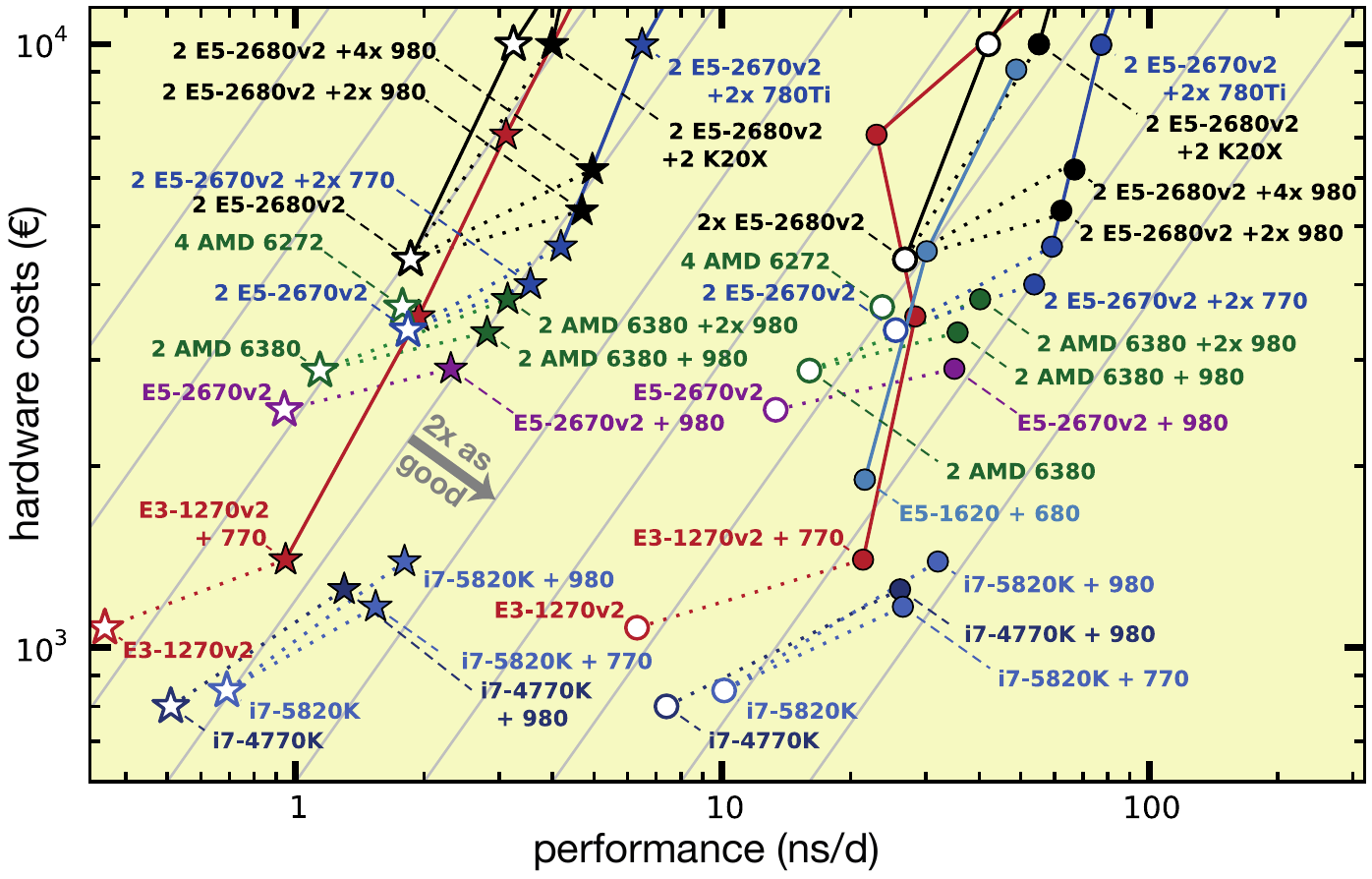}
\caption{
Benchmark performances in relation to the total hardware investment (net) for 
investments up to \PreisEuro{10000}.
\ac{MEM} (circles) and \ac{RIB} (stars)
symbols colored depending on CPU type. 
Symbols with white fill denote nodes without \ac{GPU} acceleration. 
Dotted lines connect \ac{GPU} nodes to their CPU-only counterparts.
The grey lines indicate constant performance-to-price ratio,
they are a factor of 2 apart each.
For this plot, all benchmarks not done with GCC 4.8 (see Table~\ref{tab:environment}) 
have been renormalized to 
the performance values expected for GCC 4.8, \ie plus $\approx$19\percent
for GCC 4.7 benchmarks on CPU nodes and plus $\approx$4\percent for GCC 4.7 benchmarks 
on \ac{GPU} nodes (see Table~\ref{tab:compilers}).
The costs for multiple node configurations include \PreisEuro{370}
for QDR \ac{IB} adapters (\PreisEuro{600} per FDR-14 \ac{IB} adapter) per node.
}
\label{fig:summarySingle}
\end{center}
\end{figure}
\begin{figure}[t b p]
\begin{center}
\includegraphics[width=15cm]{\pics 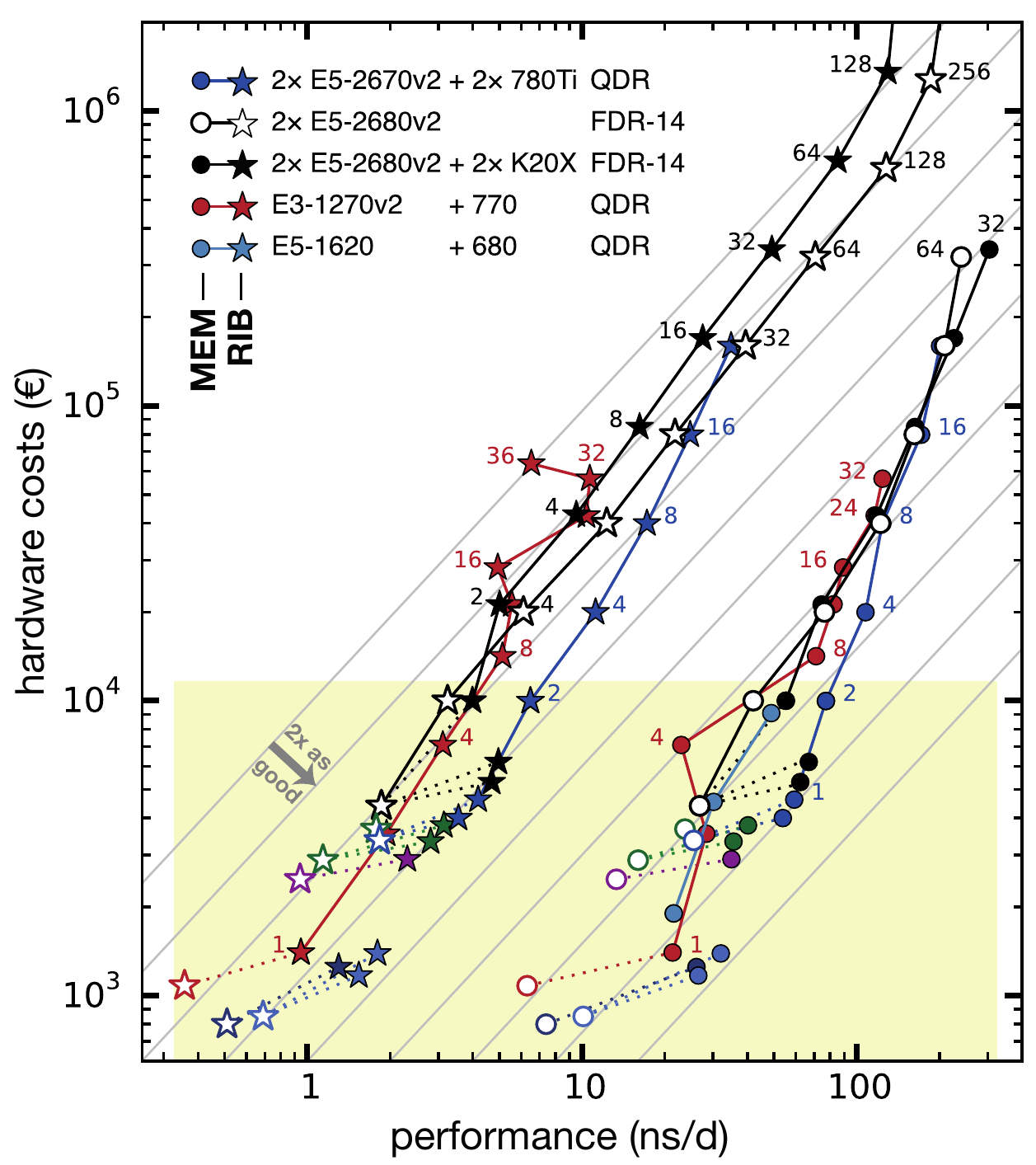}
\caption{
Same representation as in Fig.~\ref{fig:summarySingle}
(yellow box depicts section plotted there), 
now focusing on the parallel performance across multiple nodes
(the small number next to the data points indicates the number of nodes used). 
The grey lines indicate perfect scaling and constant performance-to-price ratio,
they are a factor of 2 apart each.
A number next to a data point indicates how many compute nodes were used in that benchmark.
}
\label{fig:summaryScaling}
\end{center}
\end{figure}
Figures~\ref{fig:summarySingle} and \ref{fig:summaryScaling} relate
hardware investments and performance, thus
summarizing the results in terms of our criteria
performance-to-price (C1), single-node performance (C2), and parallel performance (C3).
The grey lines indicate both perfect parallel scaling 
as well as a constant performance-to-price ratio;
configurations with better ratios appear more to the lower right.
Perhaps not unexpectedly, the highest single-node performances (C2) 
are found on the dual-CPU nodes with two or more \acp{GPU}. At the same time,
the best performance-to-price ratios (C1)
are achieved for nodes with consumer-class \acp{GPU}.
The set of single nodes with consumer \acp{GPU} 
(filled symbols in the figures)
is clearly shifted towards higher performance-to-price 
as compared to nodes without \ac{GPU} (white fill)
or with Tesla \acp{GPU}. 
Adding at least one consumer-grade \ac{GPU} to a node  
increases its performance-to-price ratio by a factor of about two, 
as seen from the dotted lines in the figures
that connect \ac{GPU} nodes with their \ac{GPU}-less counterparts.
Nodes with HPC instead of consumer \acp{GPU} (\eg\ Tesla K20X instead of GeForce GTX 980)
are however more expensive and less productive with \gromacs (black dotted lines).

Consumer PCs with an Intel Core processor and a GeForce \ac{GPU} 
in the low-cost regime at around \PreisEuro{1000} 
produce the largest amount of \ac{MD} trajectory per money spent.
However, these machines come in a desktop chassis and lack \ac{ECC} memory.
Even less expensive than the tested Core i7-4770K and i7-5830K CPUs 
would be a desktop equivalent of the E3-1270v2 system
with i7-3770 processor, which would cost about \PreisEuro{600} without \ac{GPU},
or a Haswell-based system, \eg with i5-4460 or i5-4590,
starting at less than \PreisEuro{500}.

Over the lifetime of a compute cluster,
the costs for electricity and cooling (C4) become a substantial
or even the dominating part of the total budget.
Whether or not energy costs are accounted for
therefore strongly influences what the optimal hardware
will be for a fixed budget.
Whereas the power draw of nodes with \acp{GPU} can be twice as high as without, 
their \gromacs performance is increased by an even larger factor. 
With energy costs included, 
configurations with balanced CPU/\ac{GPU} resources produce
the largest amount of \ac{MD} trajectory over their lifetime
(Tables~\ref{tab:consumption} and \ref{tab:consumption2}).

Vendors giving warranty for densely packed nodes with consumer-class \acp{GPU}
can still be difficult to find. If rack space is an issue (C5), it is possible to
mount $2\times${}Intel E26xx v2/3 processors plus up to 
four consumer \acp{GPU} in just \unit[2]{U} standard rack units.
However, servers requiring less than \unit[3]{U} that are able to host GeForce cards  
are rare and also more expensive than their \unit[3--4]{U} counterparts. 
For Tesla \acp{GPU} however, there are supported and certified solutions 
allowing for up to three \acp{GPU} and two CPUs in a \unit[1]{U} chassis.

Small tweaks to reduce hardware costs is acquiring just the minimal amount of RAM proposed by the vendor,
which is normally more than enough for \gromacs.
Also, chassis with redundant power supply adapters are more expensive but mostly unnecessary.
If a node fails for any reason, the \gromacs built-in checkpointing support ensures that
by default at most 15 minutes of trajectory production are lost and that the simulation
can easily be continued.

For parallel simulations (Figure~\ref{fig:summaryScaling}),
the performance-to-price ratio mainly depends on the parallel efficiency that is achieved.
Nodes with consumer \acp{GPU} (\eg E5-2670v2 + 2\mal{}780Ti)
connected by QDR \ac{IB} network have 
the highest performance-to-price ratios on up to about eight nodes (dark blue lines).
The highest parallel performance (or minimal time-to-solution, C3) 
for a single \ac{MD} system is recorded with the lowest latency interconnect.
This however comes at the cost of trajectories that are 2\,--\,8\mal
as expensive as on the single nodes with the best performance-to-price ratio.

Figure~\ref{fig:checklist} summarizes best practices helping to exploit
the hardware's potential with \gromacs. %
These rules of thumb for standard \ac{MD} simulations
with \ac{PME} electrostatics and Verlet cutoff scheme
hold for moderately parallel scenarios. 
When approaching the scaling limit of $\approx$ 100 atoms per core, 
a more elaborate parameter scan will be useful to find the performance optimum.
\begin{figure}
\newcommand{\mysep}{\setlength{\itemsep}{0pt}}
\renewcommand*\labelitemi{\ding{114}}
\fbox{\begin{minipage}{.95\textwidth}
\begin{center}
\begin{minipage}{.9\textwidth}
\vspace{3mm}
\fontfamily{phv}\selectfont
\small
\textbf{General hints concerning \gromacs performance}
\begin{itemize}
  \mysep 
  \item Recent GCC \textbf{compilers} $\geq 4.7$ 
        with best SIMD instruction set supported by the CPU
        produce the fastest binaries (Table~\ref{tab:compilers})
  \item In single node runs with up to 12--16 cores, $\nrank=1$ and $\nthread=\ncore$ performs best,
        on Intel even across sockets
  \item On single nodes, \textbf{thread-MPI} often offers better performance than a regular MPI library
  \item \textbf{Hyper-threading} is beneficial
       with high enough atoms/core count ($\gtrsim 2500$) and in OpenMP-only runs ($\nrank=1$)
  \item Use $\nrank \times \nthread = \ncore$ and ensure that thread \textbf{pinning} %
        works correctly (check the mdrun log output)
  \item Check \textbf{parallel efficiency} $E$ on multiple nodes
        by comparing to the single-node performance (Tables~\ref{tab:scalingAQP}--\ref{tab:scalingRIB})
  \item The \textbf{time and cycle accounting table} in 
        the \type{mdrun} log file helps discovering performance issues
\end{itemize}

\textbf{Method and algorithmic tweaks}\cite{abraham2014gromacs}
\begin{itemize}
  \mysep
  \item \textbf{Neighbor search frequency} of 20--80 is recommended, 
        $\geq 50$ values are typically useful in parallel \ac{GPU} runs (Figure~\ref{fig:nstlist})
  \item In highly parallel simulations, overhead due to \textbf{global communication} is reduced
        with lower frequencies for energy calculation, temperature and pressure coupling
  \item Consider using \textbf{h-bonds constraints} with \unit[2]{fs} time step to reduce computational and communication cost
  \item \textbf{Virtual sites} allow \unit[4--5]{fs} long time step
  \item Tweak \textbf{LINCS settings} at high parallelization
  \item Try increasing \textbf{PME order} to 5 (from the default 4) 
        with a proportionally coarser grid\cite{abraham2011optimization}
\end{itemize}

\textbf{On CPU nodes}
\begin{itemize}
  \mysep  
  \item With high core count, $\nrank=\ncore$ and $\nthread=1$ performs best and 
        \type{g\_tune\_pme} conveniently determines the optimal number of \textbf{separate \ac{PME} ranks}.
\end{itemize}

\textbf{On \ac{GPU} nodes}
\begin{itemize}
  \mysep  
  \item Try if switching \textbf{\ac{DLB}} off improves performance (Figure~\ref{fig:threads})
  \item With $\ncpu \geq 2$, using $\nrank \geq \ncpu$ often
        improves performance (Figure~\ref{fig:threads}).
  \item On Tesla K20, K40, K80, highest \textbf{application clock} rate can be used (Figure~\ref{fig:clocks})
  \item \textbf{Multi-simulation} 
        increases aggregate performance (Figure~\ref{fig:throughput})
  \item The parallel performance across many nodes can be improved by using
        a homogeneous, interleaved \ac{PME} node separation (p. \pageref{sec:homogeneous})
\end{itemize}
\end{minipage}
\end{center}
\end{minipage}}
\caption{\gromacs performance checklist.
Number of MPI ranks, \nrank;
number of OpenMP threads, \nthread;
number of CPU cores, \ncore.}
\label{fig:checklist}
\end{figure}
Unfavorable parallelization settings can reduce performance
by a factor of two even in single node runs. 
On single nodes with processors supporting \ac{HT}, for the \ac{MD} systems tested,
exploiting all hardware threads showed the best performance.
However, when scaling to higher node counts using one thread per physical core gives better performance.
On nodes with Tesla \acp{GPU}, choosing the highest supported application clock rate never hurts
\gromacs performance but will typically mean increased power consumption.
Finally, even the compiler choice can yield a 20\percent performance difference with GCC $\ge$4.7  
producing the fastest binaries.

For the researcher it does not matter from which hardware \ac{MD} trajectories originate, 
but when having to purchase the hardware it makes a substantial difference.
In all our tests, nodes with good consumer \acp{GPU} exhibit the same (or even higher)
\gromacs performance as with \ac{HPC} \acp{GPU}\,---\,at a fraction of the price. 
If one has a fixed budget,
buying nodes with expensive \ac{HPC} instead of cheap consumer \acp{GPU} means that the scientists
will have to work with just half of the data they could have had.
Consumer \acp{GPU} can be easily checked for memory integrity with 
available stress-testing tools and replaced if necessary.
As consumer-oriented hardware is not geared toward non-stop use,  
repeating these checks from time to time helps catching 
failing \ac{GPU} hardware early.
Subject to these limitations, 
nodes with consumer-class \acp{GPU} are nowadays the most economic way to
produce \ac{MD} trajectories not only with \gromacs. 
The general conclusions concerning hardware competitiveness may also have relevance for
several other \ac{MD} codes like CHARMM,\cite{CHARMM2009} LAMMPS,\cite{Brown2012449} or NAMD,\cite{phillips2005scalable} 
which like \gromacs also use \ac{GPU} acceleration in an offloading approach.

\subsection*{Acknowledgments}
Many thanks to Markus Rampp (MPG Rechenzentrum Garching) for providing help with
the \QuotMarks{Hydra} supercomputer, and to
Nicholas Leioatts, Timo Graen, Mark J. Abraham and Berk Hess for valuable suggestions on the manuscript.
This study was supported by the DFG priority programme \QuotMarks{Software for Exascale Computing}
(SPP 1648).

\subsection*{Supporting Information}
The supporting information contains examples and scripts
for optimizing \gromacs performance, as well as
the input \type{.tpr} files for the simulation systems.

\bibliography{biblio}

\end{document}

\hyphenation{off-loaded off-load off-loading}

\maketitle

\begin{abstract}
This supporting information contains \gromacs exemplary commands
to do benchmarks and performance optimization for CPU and GPU nodes.
\end{abstract}

\section{Performance optimization on CPU nodes}
\subsection*{Finding the optimal number of separate PME ranks}
To find the best performance-wise settings 
for a specific \type{.tpr} file on a node with \ncore cores without \acp{GPU}, 
it often suffices to determine the optimal number of separate \ac{PME} ranks.
The largest benefits are normally seen on nodes with a lot of cores,
though also with just a few cores, separate \ac{PME} ranks may improve performance quite a bit. 
With the following \type{bash} commands,
the optimal number of \ac{PME} ranks is determined on a node with 48 cores,
using the efficient thread-MPI implementation of \type{mdrun}.

\begin{lstlisting}[caption={}]
# Add the GROMACS programs to your path:
source /path/to/your/gromacs/bin/GMXRC

# Let g_tune_pme know how the thread-enabled mdrun is called: 
export MDRUN=$( which mdrun_threads )

# Invoke the bunch of tuning runs with your .tpr file:
g_tune_pme -ntmpi 48 -s in.tpr -cpt 1440
\end{lstlisting}
The optional \type{-cpt 1440} argument sets the interval of checkpoint writing to 24 hours,
to suppress unnecessary checkpoints during benchmarking.
Please look up other helpful tuning options
with \type{g\_tune\_pme -h} (for \gromacs 4.6), or \type{gmx help tune\_pme} (for 5.x).

If you need to use a regular MPI library, adapt the following commands
to your needs.
\begin{lstlisting}[caption={}]
# Point to the GROMACS bin directory, let g_tune_pme know how the 
# MPI-enabled mdrun is called and what command is needed 
# to start parallel runs:
source /path/to/your/gromacs/bin/GMXRC
export MPIRUN="/path/to/your/mpi/mpirun -machinefile hosts"
export MDRUN=$( which mdrun_mpi )
g_tune_pme -np 48 -s in.tpr
\end{lstlisting}
Tuning will produce a table with all successfully tested settings 
and the resulting performances.
For parallel runs across multiple nodes 
a machine file may be needed depending on the MPI distribution 
and\,/\,or queuing system in use.

Here is an example using IBM's LoadLeveler, 
which requires that parallel jobs are started using \type{poe} (not mpirun).
Moreover, the number of ranks has to be specified via queue parameters exclusively,
and not with \type{-np} as for other MPI frameworks;
this is taken care of by \type{-npstring none}.
In this example intended for 20-core nodes, 
altogether 640 MPI processes are started on 64 nodes (10 MPI tasks per node),
each using two OpenMP threads, thus making use of 20 cores per node.

\begin{lstlisting}[caption={}]
# @ shell=/bin/bash
# @ job_type = parallel
# @ node_usage= not_shared
# @ node = 64                        # request 64 nodes
# @ tasks_per_node = 10              # start 10 MPI processes per node
# @ resources = ConsumableCpus(2)    # use 2 threads per MPI process
# @ queue
module load gromacs/4.6.7            # e.g.
export MDRUN=/path/to/mdrun_mpi

# This is important so that g_tune_pme knows how to start MPI processes:
export MPIRUN=poe

# Do the tuning runs! Note that -ntomp 2 is not strictly needed here,
# since already specified by the ConsumableCpus(2) line above
g_tune_pme -np 640 -npstring none -ntomp 2 -s in.tpr 
\end{lstlisting}

If one does not want to perform extensive run parameter scans to find the performance optimum,
the \type{mdrun} log file \type{md.log} still provides useful information on
what parameters to change to approach the performance optimum.
For example, running the \ac{MEM} benchmark on the 40 cores of a 2\mal E5-2680 node 
using 28 \ac{PP} plus 12 \ac{PME} ranks, results in a performance of \unit[20]{\nsday}.
Near the end of the log file, the work balance between the \ac{PME} mesh part
(computed on the \ac{PME} ranks) and the rest of the force calculation
(computed on the \ac{PP} ranks) is listed, 
which turns out to be 0.625 in this case.

\begin{lstlisting}[caption={}]
 Average PME mesh/force load: 0.625
 Part of the total run time spent waiting due to PP/PME imbalance: 8.3 %

NOTE: 8.3 % performance was lost because the PME nodes
      had less work to do than the PP nodes.
      You might want to decrease the number of PME nodes
      or decrease the cut-off and the grid spacing.
\end{lstlisting}

As noted, this is unfavourable as the \ac{PME} ranks are idle a significant amount of time
each step to wait for the \ac{PP} computations to finish.
It should be checked whether performance is increased
with a lower number of \ac{PME} ranks. 
Using 10 \ac{PME} ranks indeed yields a performance of about \unit[26]{\nsday}
and a balance very near to 1.0 between \ac{PME} and \ac{PP} computation.

\begin{lstlisting}[caption={}]
 Average PME mesh/force load: 1.011
 Part of the total run time spent waiting due to PP/PME imbalance: 0.7 %
\end{lstlisting}

\section{Performance optimization on \ac{GPU} nodes}
\label{sec:gpu}

For optimal performance on \ac{GPU} nodes 
it is necessary to achieve a balanced load distribution between \ac{GPU} and CPU.
How well the load is balanced is obtained from
the GPU\,/\,CPU force evaluation time
listed at the end of the \type{md.log} output file.
\begin{lstlisting}[caption={}]
 Force evaluation time GPU/CPU: 5.673 ms/8.301 ms = 0.683
For optimal performance this ratio should be close to 1!

NOTE: The GPU has >25% less load than the CPU. This imbalance causes
      performance loss.
\end{lstlisting}
Here, the automatic tuning of Coulomb cutoff and \ac{PME} grid spacing was deactivated,
resulting in a performance loss due to a too small Coulomb cutoff and too large \ac{PME} grid,
which is clearly disadvantageous.

On the other hand, a very large Coulomb cutoff resulting from automatic \ac{GPU}\,/\,CPU
load balancing can be an indicator for a suboptimal hardware configuration.
Near the end of the \type{md.log} file,
the values of the Coulomb cutoff and \ac{PME} grid spacing are listed.
The first line provides the input values from the \type{.tpr} file,
the following line the optimized values after \ac{GPU}\,/\,CPU load balancing, 
here for the example of the 4-\ac{GPU} benchmark from Table~\ref{tab:example}
in the main text.
\begin{lstlisting}[caption={}]
 PP/PME load balancing changed the cut-off and PME settings:
           particle-particle                    PME
            rcoulomb  rlist            grid      spacing   1/beta
   initial  1.000 nm  1.012 nm     240 240 240   0.130 nm  0.289 nm
   final    1.607 nm  1.619 nm     144 144 144   0.217 nm  0.465 nm
 cost-ratio           4.10             0.22
\end{lstlisting}
The fact that the Coulomb cutoff is increased by 60\percent 
(or that the \acp{GPU} get 4.1 times as much computational load as a result of tuning)
indicates a lack of CPU compute power relative to the available \ac{GPU} power on this node,
resulting in decreased performance\,/\,Watt.

\subsection*{Finding the optimal number of threads per rank}
On \ac{GPU} nodes with more than $\approx 8$ cores 
and\,/\,or with more than one \ac{GPU},
you may want to derive the perfect combination of MPI ranks and OpenMP threads
(see Figure~\ref{fig:threads} in the main article).
When the number of ranks on a node does not equal the number of \acp{GPU},
a string has to be provided to \type{mdrun} 
with the proper mapping of \acp{GPU} to ranks. 
\Eg, to use 6 \ac{PP} ranks with 2 \acp{GPU}, 
the following command assigns \ac{GPU}~0 to the first three ranks and \ac{GPU}~1 to the last three.

\begin{lstlisting}[caption={}]
mdrun -ntmpi 6 -gpu_id 000111 -s in.tpr
\end{lstlisting}

The following \type{bash} function constructs a 
\type{-gpu\_id} string based on the number of \ac{PP} ranks and \acp{GPU}
and is useful when doing a ranks versus threads parameter scan.
\lstinputlisting[basicstyle=\alterLineSpreadStyle]{\listingsDir constructGpuString.sh}

Here is an example for 4 \acp{GPU} node, for
which the function will create the string \QuotMarks{\type{0001122233}} 
when requesting 10 ranks per node. 
On a 40-core node \type{mdrun} will automatically fill up all available cores 
by choosing 4 threads per rank. 

\begin{lstlisting}[caption={}]
# Example:
USEGPUIDS="0123456789"  # only GPU id's from this list will be used
NGPU_PER_NODE=4
NRANK_PER_NODE=10

# Construct a string that maps ranks to GPUs
GPUSTR=$( func.getGpuString $NGPU_PER_NODE $NRANK_PER_NODE $USEGPUIDS )

# ... and use it!
mdrun_threads -s in.tpr -ntmpi $NRANK_PER_NODE -gpu_id $GPUSTR
\end{lstlisting}

A parameter scan for the optimal division of cores in ranks and threads 
$\ncore = \nrank \mal \nthread$ could be done using the following script.
Note that \gromacs requires at least one rank per \ac{GPU}.  

\begin{lstlisting}[caption={}]
CORES=40
NGPU_PER_HOST=2
USEGPUIDS="0123456789"
DIR=$( pwd )

# list with all possible numbers of thread-MPI ranks to check.
# Threads will be started automatically by mdrun to fill up all cores
RANKLIST="40 20 10 8 5 4 2"

for NRANK in $RANKLIST ; do
  GPUSTR=$( func.getGpuString $NGPU_PER_HOST $NRANK $USEGPUIDS )
  for DLB in "no" "yes" ; do
    for RUN in ntmpi${NRANK}_dlb${DLB} ; do
      mkdir "$DIR"/run_$RUN
      cd "$DIR"/run$RUN
      mdrun -ntmpi $NRANK -npme 0 -s in.tpr -dlb $DLB -gpu_id $GPUSTR
    done
  done
done
\end{lstlisting}

\subsection*{Separate PME ranks on \ac{GPU} nodes}
If running on multiple \ac{GPU} nodes in parallel,
one might want to assign half of the ranks per node as separate \ac{PME} ranks.
This can be achieved with settings similar to the following,
which gives an example using LoadLeveler on 32 nodes with 2 \acp{GPU} each.
Since the \type{mdrun} default is an \emph{interleaved} rank assignment,
each node will get assigned two \ac{PP} and two \ac{PME} ranks.
With the \type{-gpu\_id 01} string,
these two \ac{PP} ranks on a node are properly assigned to the two \acp{GPU}.

\begin{lstlisting}[caption={}]
# @ tasks_per_node = 4      # 4 ranks per node (2 for PP, 2 for PME)
# @ node = 32               # 32 nodes, i.e. 128 MPI ranks altogether
# @ resources = ConsumableCpus(5)  # use 5 threads per rank for 20 cores

# Use half of the 128 ranks for PME:
poe mdrun_mpi -npme 64 -s in.tpr -gpu_id 01
\end{lstlisting}

This example is similar to the above one, 
now using the more common OpenMPI~/ IntelMPI~/ MPICH style process startup.
Additionally, here the 20 cores are distributed non-evenly across \ac{PME} and
\ac{PP} ranks to fine-tune the \ac{PP}-\ac{PME} compute power.
\begin{lstlisting}[caption={}]
mpirun -np 128 mdrun_mpi -npme 64 -ntomp 4 -ntomp_pme 6 -gpu_id 01 ...
\end{lstlisting}

\subsection*{Optimizing throughput with multi simulations}
\label{sec:multi}
$M$ independent simulations can be started using 
either the \type{-multi <M>} or the \type{-multidir <dirA dirB dirC ...>} option
of an MPI-enabled \type{mdrun}, which automatically takes care of proper process placement and
pinning. Note that in case of more than one \ac{GPU} on a node a proper
\type{-gpu\_id} string has to be provided, mapping ranks to \acp{GPU}, 
as given in the following examples for two \acp{GPU} and eight replicas.
This example uses the \type{-multi} keyword, 
in which case the individual \type{.tpr} input files 
need to have the replica number attached to the base file name, 
\ie here \type{repl0.tpr}, \ldots, \type{repl3.tpr}.  

\begin{lstlisting}[caption={}]
mpirun -np 4 mdrun_mpi -multi 4 -gpu_id 0011 -s repl.tpr
\end{lstlisting}

Alternatively, with the \type{-multidir} mechanism, 
using the run directories \type{replA}, \ldots \type{replD}, each containing
an input file \type{topol.tpr}, the command is the following.
\begin{lstlisting}[caption={}]
mpirun -np 4 mdrun_mpi -multidir replA replB replC replD -gpu_id 0011
\end{lstlisting}

\type{mdrun} will automatically determine the number of cores per node \ncore 
and assign $\nthread = \ncore / M $ threads to each replica.
As another option, the number of threads can be controlled with the optional
command line parameters \type{-ntomp} (threads per replica) or \type{-nt} (total threads),
which is useful if performance with and without hyper-threading shall be compared.

\begin{table}[tb]
\caption{Enhanced parallel efficiency by running multi-simulations across several nodes,
here for the \ac{RIB} benchmark 
on 2\mal{}E5-2680v2 nodes with 2\mal{}K20X \acp{GPU} 
(see lowermost part of Table~\ref{tab:scalingRIB} in the main text).
$P_1$ and $E_1$ give the performance and parallel efficiency 
for a single replica densely packed on the available nodes, 
whereas $P_4$ and $E_4$ are recorded for an interleaved placement 
of 4 replicas on 4 times the number of nodes.}
\label{tab:multiRIB}
\STautoround{2}
\begin{center}
\begin{spreadtab}{{tabular}{r N{5}{3} l P l r r c }} \toprule
@nodes\,/  &@\mult{$P_1$}      & @\mult{$E_1$}            &@\mult{$P_4$}      & @\mult{$E_4$}            &@\multicolumn{2}{c}{per replica} &         \\
@replica   &@\mult{(\nsday)}   &                          &@\mult{(\nsday)}   &                          &@\mult{\nrank} &@\mult{\nthread} &         \\ \midrule
    1      & round(  3.986, 2) & \STcopy{v}{b3/(!a3*b!3)} & round(  3.747, 2) & \STcopy{v}{d3/(!a3*b!3)} &   20          &    2            &@ \hy    \\
    2      & round(  5.014, 2) &                          & round(  6.563, 2) &                          &   80          &    1            &@ \hy    \\
    4      & round(  9.526, 2) &                          & round( 12.091, 1) &                          &  160          &    1            &@ \hy    \\
    8      & round( 16.247, 1) &                          & round( 20.774, 1) &                          &  160          &    1            &         \\
   16      & round( 27.475, 1) &                          & round( 33.568, 1) &                          &  160          &    4            &@ \hy    \\
   32      & round( 49.095, 1) &                          & round( 55.480, 1) &                          &  320          &    2            &         \\
   64      & round( 85.332, 1) &                          & round( 94.814, 1) &                          &  640          &    2            &         \\
%   128    & round(129.667, 1) &                          &                   &@                         \\
%  256    & round(139.535, 1) &                          &                   &@                         \\
\bottomrule
\end{spreadtab}
\end{center}
\end{table}
The multi-simulation technique can also be used to enhance the performance across several nodes.
This requires a fast enough interconnect, as \eg\ FDR-14 Infiniband in our test case, see Table~\ref{tab:multiRIB}.
Instead of a \emph{dense} placement of replicas, where each replica is spread across
as few nodes as possible, each node gets one or more \ac{DD} domains of each replica.
As a result, the replicas are spread across all available nodes in an \emph{interleaved} fashion.
Like in a single-node multi-simulation, the individual replicas may run (slightly) out of sync,
allowing for better \ac{GPU} utilization and raised performance.
In addition, as the individual replicas perform their communication at different times,
less CPU cores compete at the same time for the available network resources as compared to a dense placement.
This setup is especially useful for larger \ac{MD} systems, where running multiple copies on a 
single node would be too slow.
\Eg, when running an ensemble of 4 \ac{RIB} replicas using 16 nodes one could start 4 completely independent
simulations, each using 4 nodes, which would yield a performance of \unit[9.53]{\nsday}
per replica (Table~\ref{tab:multiRIB}). 
With multi-simulations, the 4 replicas are distributed across
the 16 nodes in an interleaved way, which yields \unit[12.1]{\nsday} per replica
and thus a \unit[27]{\percent} increased performance.

\subsection*{Fixing the \ac{GPU} fan speed}
The probability of \ac{GPU} throttling of consumer \acp{GPU} can be diminished 
by setting the fan speed to a value higher than the default
(see Section~\ref{sec:applicationClocks} in the main document).
The \ac{GPU} fan speed can not be set using the \type{nvidia-smi}; instead, the graphical 
utility and a running X server is required. First, the \type{xorg.conf} config file 
should be modified to include the \QuotMarks{\type{coolbits}} option:
\begin{lstlisting}[caption={}]
Section "ServerLayout"
        Identifier "dual"
        Screen 0 "Screen0"
        Screen 1 "Screen1" RightOf "Screen0"
EndSection

Section "Device"
        Identifier    "nvidia0"
        Driver        "nvidia"
        VendorName    "NVIDIA"
        BoardName     "GeForce GTX TITAN"
        Option        "UseDisplayDevice" "none"
        Option        "Coolbits" "4"
        BusID         "PCI:2:0:0"
EndSection

Section "Device"
[...]
EndSection

Section "Screen"
        Identifier    "Screen0"
        Device        "nvidia0"
EndSection

Section "Screen"
[...]
EndSection
\end{lstlisting}

Then, start the X server and set the fan speeds:
\begin{lstlisting}[caption={}]
export DISPLAY=:0

/usr/bin/X $DISPLAY -nolisten tcp vt7 -novtswitch &

nvidia-settings -a [gpu:0]/GPUFanControlState=1 \
   -a [fan:0]/GPUCurrentFanSpeed=80

nvidia-settings -a [gpu:1]/GPUFanControlState=1 \
   -a [fan:1]/GPUCurrentFanSpeed=80
\end{lstlisting}

The fan speed settings can be checked with \type{nvidia-smi}:
\begin{lstlisting}[caption={}]
+------------------------------------------------------+
| NVIDIA-SMI 346.47     Driver Version: 346.47         |
|-------------------------------+----------------------+
| GPU  Name        Persistence-M| Bus-Id        Disp.A |
| Fan  Temp  Perf  Pwr:Usage/Cap|         Memory-Usage |
|===============================+======================+
|   0  GeForce GTX TITAN   On   | 0000:02:00.0     Off |
|  80%  69C    P8  205W /  250W |    549MiB /  6143MiB |
+-------------------------------+----------------------+
|   1  Quadro M6000        On   | 0000:03:00.0     Off |
|  80%  66C    P0  203W /  250W |    533MiB / 12287MiB |
+-------------------------------+----------------------+
\end{lstlisting}